\documentclass[journal]{IEEEtran} 
\usepackage[draft]{graphicx}
\usepackage{graphicx}
\usepackage{amsmath}
\usepackage{amssymb}
\usepackage{booktabs}
\usepackage{enumerate}
\usepackage[ruled]{algorithm2e}
\usepackage{algorithmic}
\usepackage{amsmath}
\usepackage{graphics}
\usepackage{epsfig}
\usepackage{subfigure}
\usepackage{mmap} 
\usepackage{multirow}

\usepackage{CJK}
\usepackage{mathtools}
\usepackage{bbding}
\usepackage{amsmath}
\usepackage{amssymb}
\usepackage{ntheorem}
\usepackage{graphicx}
\usepackage{xcolor}

\psfull

\begin{document}
\title{{Securing Reconfigurable Intelligent Surface-Aided Cell-Free Networks}}
\author{Wanming Hao,~\IEEEmembership{Member,~IEEE,} Junjie Li, Gangcan Sun, Ming Zeng,~\IEEEmembership{Member,~IEEE,} Octavia A. Dobre,~\IEEEmembership{Fellow,~IEEE}				
		 \thanks{W. Hao is with the Henan Institute of Advanced Technology,
        and also with the School of Information Engineering,  Zhengzhou University, Zhengzhou 450001, China. (e-mail: iewmhao@zzu.edu.cn)}
		\thanks{J. Li is with the Henan Institute of Advanced Technology, Zhengzhou University, Zhengzhou 450001, China. (e-mail: 202012592017620@gs.zzu. edu.cn)}
        \thanks{G. Sun is with the School of Information Engineering, and Institute of Industrial Technology, Zhengzhou University, Zhengzhou 450001, China. (Email: iegcsun@zzu.edu.cn)}  
        \thanks{M. Zeng is with the Department of Electrical Engineering and Computer Engineering, Universite Laval, Quebec, G1V 0A6, Canada. (E-mail: ming.zeng@gel.ulaval.ca)}
        \thanks{O. A. Dobre is with the Faculty of Engineering and Applied Science, Memorial University, St. Johns, NL A1B 3X5, Canada. (E-mail:  odobre@mun.ca)}}

\maketitle

\begin{abstract}
In this paper, we investigate the physical layer security in the reconfigurable intelligent surface (RIS)-aided cell-free networks.  A maximum weighted sum secrecy rate problem is formulated by jointly optimizing the active beamforming (BF) at the base stations and passive BF at the RISs. To handle this non-trivial problem, we adopt the alternating optimization to decouple the original problem into two sub-ones, which are solved using the semidefinite relaxation and continuous convex approximation theory. To decrease the complexity for obtaining overall channel state information (CSI), we extend the proposed framework to the case that only requires part of the RIS' CSI. This is achieved via deliberately discarding the RIS that has a small contribution to the user's secrecy rate. Based on this, we formulate a mixed integer non-linear programming problem, and the linear conic relaxation is used to obtained the solutions. Finally, the simulation results show that the proposed schemes can obtain a higher secrecy rate than the existing ones.
\end{abstract}


\IEEEpeerreviewmaketitle
\setlength{\baselineskip}{1\baselineskip}
\newtheorem{definition}{Definition}
\newtheorem{fact}{Fact}
\newtheorem{assumption}{Assumption}
\newtheorem{theorem}{Theorem}
\newtheorem{lemma}{Lemma}
\newtheorem{corollary}{Corollary}
\newtheorem{proposition}{Proposition}
\newtheorem{example}{Example}
\newtheorem{remark}{Remark}

\section{Introduction}
With the rapid development of the smart wireless applications, the demand of the wireless data throughput has been speedily increasing [1]. To satisfy the above requirement, more base stations (BSs) can be deployed to improve the service capacity [2], [3]. However, the serious interference will occur with traditional cellular network structure, especially for the cell-edge users [4]. To address this, the user-centric cell-free network structure has been developed, where multiple BSs cooperatively serve the users [5], [6]. In this way, there is no cell boundary, thereby effectively integrating the inter-cell interference and improving the system throughput as well as coverage probability [7], [8].

However, the ultra-dense BS deployment inevitably leads to huge energy consumption. Recently, the reconfigurable intelligent surface (RIS) technology has been proposed as an energy efficient solution, where the RIS is composed of a large number of low-cost nearly passive reflective elements with controller phase shifts [9], [10]. The RIS can reflect the received signal to the desired direction by adjusting the reflective elements, which effectively improves the strength of the received signals [11]. Compared with the traditional relay, there is no noise amplification and self interference in RIS, moreover, RIS can greatly decrease the energy consumption and is easy to deploy [12]. Therefore, it is a promising scheme to replace some BSs with RISs in cell-free networks, such that the high rate and low energy consumption can be obtained simultaneously.

Due to the inherent broadcasting characteristics of wireless signals, any user within the coverage range may receive them and could launch various attacks (eavesdropping, monitoring, etc.) [13], especially for the cell-free networks. Nowadays, the physical layer security (PLS) technique has been adopted to deal with the security issue [14], [15]. Interestingly, the RIS has also showed a good performance in improving the system security, through increasing the signal strength of legitimate users and weakening that of illegal users [16]. Motivated by this potential advantage of RIS, we will study the PLS in RIS-based cell-free networks in this paper.
\subsection{Related Works}
Recently, there have been several innovative works considering the security performance of the cell-free networks [17-18]. Specifically, [17] studied the achievable secrecy rates and compared that with the co-located massive MIMO networks, where the active eavesdropper (Eve) was assumed. The authors in [18] considered the spatially correlated Rayleigh fading channels with active Eve, and proposed a distributed conjugate beamforming (BF) technique to obtain the minimum mean square error (MMSE). Furthermore, their research verified the correlation of ergodic secrecy rate with active Eve and channel fading conditions.

Meanwhile, RIS-aided secure issues have also received great attention recently [19-23]. Specifically, in [19], the authors considered a RIS-aided multiple-input single-output (MISO) system, in which an Eve eavesdrops the confidential message from the access point (AP) to a specific user. The authors then proposed a joint AP active BF and RIS passive BF scheme to maximize the secrecy rate of all users. Under the same system model, the authors in [20] considered AP-RIS as a rank one channel, and proposed a projection gradient descent algorithm to maximize the secrecy rate. To further enhance the PLS performance, artificial noise was introduced in [21] to deliberately destroy Eve's channel. In [22], RIS-aided multi-user two-way communication secure system was considered, where one user signal was used to interfere with Eve wiretapping other users' information. Considering the practical dynamic environment, a novel deep reinforcement learning (DRL)-based secure BF design method was proposed  to guarantee the security by dynamically adjusting to the beam according the environment [23].
\subsection{Our Contributions}
However, most of the recent related works only consider the single BS in RIS communications, without touching the cooperative secure BF problem.
 It is of significance to investigate RIS-aided cell-free networks, in order to further boost the network capacity or lower energy consumption. Nonetheless, the joint BF design at the BSs and RISs is much more challenging for the multi-BS and multi-RIS systems. In this paper, we aim to address this by studying the secrecy transmission problem in a RIS-based cell-free networks. The main contributions are summarized as follows:
\begin{itemize}
 \item[$\bullet$]
 We construct a RIS-based cell-free network under the presence of an Eve, where all RISs and BSs rely on the central processing unit (CPU) to cooperatively serve all users at the same time. Based on this, we formulate a maximum weight sum secrecy rate (WSSR) problem by jointly optimizing the active cooperating BF at BSs and passive cooperating BF at RISs.
 \item[$\bullet$]
 To address the formulated non-trivial WSSR maximization problem, we propose a joint optimization framework based on all known channel state information (CSI). Specifically, we decompose the original problem into two sub-problems, and then each sub-problem is transformed into a solvable convex problem through semi-definite relaxation (SDR) and successive convex approximation (SCA) techniques. Finally, the solution of each sub-problem is updated though the proposed alternating optimization (AO) method until convergence.
 \item[$\bullet$]
 Due to the high-dimensional channel maxtrix of RISs, the channel estimation is a great challenge. In addition, it is also unrealistic to continuously acquire all RIS CSI at each small coherence time. To overcome this issue, we propose an extension of the joint optimization framework based on assigning RIS. Specifically, we first obtain all CSI once at the beginning of each large coherence time, and then introduce the matching factor of user and RIS to transform the original optimization problem into a mixed-integer non-linear programming (MINP) one. Next, the upper bound of the objective function is obtained by applying the SCA method, while the linear conic relaxation (LCR) method is applied to relax zero-one constraint for obtaining a suitable matching factor. Based on the matching factor, the maximum WSSR is obtained by jointly optimizing the BF of the BSs and the RISs in each small coherence time.
\end{itemize}

The rest of this paper is organized as follows:  Section II proposes a downlink RIS-aided cell-free secure system model, and formulates a joint optimization of the maximization WSSR problem. A joint optimization framework is proposed in Section III, and the extension framework of the assigning RIS scheme is proposed in Section IV. Section V gives a supplement to the framework. Section VII provides simulation results to verify the security performance of the above frameworks. Section VIII concludes the paper.

\emph{Notation:}
In this paper, boldface letters indicate matrices or vectors. $\mathbb{C}^{M\times N}$ denotes a complex-valued matrix of size $M\times N$. $\mathbf{A}^T$, $\mathbf{A}^H$, Tr$(\mathbf{A})$ respectively denote transpose, conjugate transpose, and trace of matrix $\mathbf{A}$. For a complex value $x$, $\angle \left( x \right)$ represents its phase, $\left| x \right|$ represents its modulus, and $\Re(x)$ is its real part. For a vector $\mathbf{a}$, ${\left[ {\bf{a}} \right]_i}$ represents its $i$-th element, $\left\| {\bf{a}} \right\|$ represents its Euclidean norm, and ${\rm{diag}}\left( x \right)$ represents the diagonal operation. For a matrix ${\bf{A}}$, ${{\bf{A}}_{i,j}}$ is the element in its $i$-th row and $j$-th column, and ${\bf{A}}\succeq 0$ represents the semi-definite matrix. ${{\bf{1}}_L}$ is a vector of length $L$ with all elements 1, ${{\bf{0}}_{L \times M}}$ represents an all-zero element matrix of size $L \times M$, ln$(\cdot)$ denotes the natural logarithm, and ${\nabla _x}$ is the gradient of $x$.

\begin{figure}[t]
	\centering
	\includegraphics[scale = 0.4]{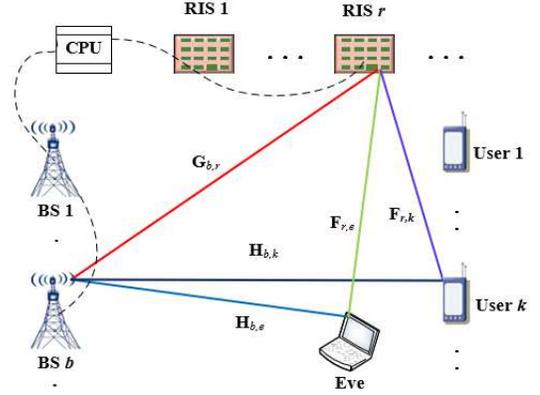}
	\caption{The IRS-aided cell-free network MISO system model.}
	\label{fig:system}
\end{figure}
\section{System Model and Problem Formulation}
In this section, we first describe the PLS-based RIS-aided cell-free networks, and then formulate the WSSR problem under several constraints.
\subsection{System Model}
As shown in Fig. 1, in the PLS-based RIS-aided cell-free network we assume that there are $B$ BSs, $R$ RISs, and $K$ users, which are randomly deployed in a given area. The BSs are connected to the CPU through backhaul links [24] and RISs are wired or wirelessly connected by the CPU or BSs [25]. For simplification, we assume that there exits a single Eve for  wiretapping all users' information. In addition, the BSs and RIS are equipped with $M$ and $N$ antennas, respectively, while the users and Eve are equipped with single antenna, respectively. Let ${\cal B} = \left\{ {1, \ldots ,B} \right\}$, ${\cal R} = \left\{ {1, \ldots ,R} \right\}$, ${\cal N} = \left\{ {1, \ldots ,N} \right\}$ and ${\cal K} = \left\{ {1, \ldots ,K} \right\}$ denote the index of BSs, RISs, RIS elements and users, respectively.

Without loss of generality, we assume that there exist both direct links (from BSs to users/Eve) and reflect links (from BSs to RISs  and from RISs to users/Eve).  The direct channels from the $b$-th BS to the $k$-user and Eve are denoted as ${\mathbf{H}}_{b,k} \in {{\mathbb{C}}^{M \times {1}}}$ and ${\mathbf{H}}_{b,e} \in {{\mathbb{C}}^{M \times {1}}}$, respectively.  ${\mathbf{G}}_{b,r} \in {{\mathbb{C}}^{N \times {M}}}$, ${\mathbf{F}}_{r,k} \in {{\mathbb{C}}^{N \times {1}}}$ and ${\mathbf{F}}_{r,e} \in {{\mathbb{C}}^{N \times {1}}}$ respectively represent link channels from the $b$-th BS to the $r$-th RIS, from the $r$-th RIS to the $k$-th user, and from the $r$-th RIS to the Eve. To characterize the performance of the considered RIS-aided cell-free secrecy network, all CSI is assumed available via the advanced channel estimation techniques. Here, we assume that all legitimate users and Eve do not collude [26].

The reflection phase matrix of the $r$-th IRS is represented by ${{\bf{\Theta }}_r} = {\rm{diag(}}{\theta _{r{\rm{,}}1}},\ldots, {\theta _{r{\rm{,}}N}}) \in {{\mathbb{C}}^{N \times {N}}}$, where ${\theta _{r{\rm{,}}n}}$ is the reflection phase of the $n$-th reflection element of the $r$-th IRS. Each reflecting unit of RISs adjusts the phase of the incident signal to suppress the Eve's reception and enhance the users' reception. Consider that the signal power reflected by RISs twice or more is much lower than the channel reflected by RISs once [27], [28], they are ignored here. We further assume ideal RIS here, i.e., each element ${\theta _{r{\rm{,}}n}}$ can be independently and continuously regulated [29], i.e.,
\begin{equation}\label{eq:1}
{\cal F}\!=\!\left\{ {{\theta _{r,n}}{\rm{|}}{\theta _{r,n}}\!=\! {\kappa _{r,n}}{e^{j{\varphi _{r,n}}}},{\kappa _{r,n}} \!\in\! \left[ {0,1} \right]\!,\! {\varphi _{r,n}} \!\in\! \left[ {0\!,\!\left. {2\pi } \right)} \right.} \right\},
\end{equation}
where $\forall r \in {\cal R}$, $\forall n \in {\cal N}$. To maximize the utility of the RISs, suppose ${\kappa _{r,n}}=1$. Note that it is unrealistic to achieve continuous adjustment of the reflection element phase in practice [30], [31], so we will refer to the low-resolution discrete phase shift in Section V.

Let $s_k$ denote the transmit signal for the $k$-user, where ${\mathbb{E}}\left\{ {{{\left| {{s_k}} \right|}^2}} \right\} = 1$. The transmit signals by the $b$-th BS can be expressed as
\begin{equation}\label{eq:2}
{{\bf{x}}_b} = \sum\limits_{k = 1}^K {{{\bf{w}}_{b,k}}{s_k}},
\end{equation}
where ${{\bf{w}}_{b,k}} \in {{\mathbb{C}}^{M \times {1}}}$ represents the precoding vector for the $k$-user. Thus, received signal of the $k$-th user can be expressed as

\noindent
\begin{equation}\label{eq:3}
\begin{aligned}
{y_k} =& \sum\limits_{b = 1}^B \left( {\bf{H}}_{b,k}^H + \sum\limits_{r = 1}^R {{\bf{F}}_{r{\rm{,}}k}^H} {\bf{\Theta }}_r^H{{\bf{G}}_{b,r}}\right){{\bf{x}}_b}+ {n_k}\\
 =& {\kern 1pt} \sum\limits_{b = 1}^B \left( {\bf{H}}_{b,k}^H + \sum\limits_{r = 1}^R {{\bf{F}}_{r,k}^H} {\bf{\Theta }}_r^H{{\bf{G}}_{b,r}}\right){{\bf{w}}_{b,k}}{{{s}}_{k}} + \\
&{\rm{ }}\sum\limits_{b = 1}^B {\sum\limits_{j\!=\!1,j\!\ne\!k}^K  }\left( {\bf{H}}_{b,k}^H\!+\!\sum\limits_{r\!=\!1}^R {{\bf{F}}_{r,k}^H} {\bf{\Theta }}_r^H{{\bf{G}}_{b,r}}\right){{\bf{w}}_{b,j}}{{{s}}_{j}}\!+\! {n_k},
\end{aligned}
\end{equation}
where ${n_k}$ is the additive white Gaussian noise (AWGN) with zero mean and variance $\sigma _k^2$. The first term on the right hand-side of the second equal sign represents the desired signal for the $k$-th user, while the second term represents the interference from all other users' signals. At the same time, the $k$-th user's information  wiretapped by the Eve can be written as
\begin{equation}\label{eq:4}
\begin{split}
y_k^e =& \sum\limits_{b = 1}^B \left( {\bf{H}}_{b,e}^H + \sum\limits_{r = 1}^R {{\bf{F}}_{r,e,k}^H} {\bf{\Theta }}_r^H{{\bf{G}}_{b,r}}\right){{\bf{x}}_b}+ {n_e} \\
 =& {\kern 1pt} \sum\limits_{b = 1}^B \left( {\bf{H}}_{b,e}^H + \sum\limits_{r = 1}^R {{\bf{F}}_{r,e,k}^H} {\bf{\Theta }}_r^H{{\bf{G}}_{b,r}}\right){{\bf{w}}_{b,k}}{{{s}}_{k}} + \\
&{\rm{ }}\sum\limits_{b \!=\! 1}^B {\sum\limits_{j\! =\! 1,j \!\ne\! k}^K  }\left( {\bf{H}}_{b,e}^H \!+\! \sum\limits_{r \!=\! 1}^R {{\bf{F}}_{r,e,k}^H} {\bf{\Theta }}_r^H{{\bf{G}}_{b,r}}\right){{\bf{w}}_{b,j}}{{{s}}_{j}} \!+\! {n_e},
\end{split}
\end{equation}
where ${n_e}$ is AWGN with zero mean and variance ${\sigma_e^2}$. In addition, we simplify the following formulation as
\begin{equation}\label{eq:5}
\begin{split}
&{\rm{     }}\sum\limits_{b = 1}^B \left( {\bf{H}}_{b,k}^H + \sum\limits_{r = 1}^R {{\bf{F}}_{r,k}^H} {\bf{\Theta }}_r^H{{\bf{G}}_{b,r}}\right){{\bf{w}}_{b,k}}\\
\mathop  = \limits^{(a)} &{\rm{ }}\sum\limits_{b = 1}^B \left( {\bf{H}}_{b,k}^H + \sum\limits_{r = 1}^R {\boldsymbol{\theta }_r^H}{\rm{diag}}\left({{\bf{F}}_{r,k}^H} \right){{\bf{G}}_{b,r}}\right){{\bf{w}}_{b,k}}\\
\mathop  = \limits^{(b)} &{\rm{ }}\sum\limits_{b = 1}^B \left( {\bf{H}}_{b,k}^H + {\boldsymbol{\theta} ^H}{\rm{diag}}\left({\bf{F}}_k^H\right){{\bf{G}}_{b}}\right){{\bf{w}}_{b,k}}\\
\mathop  = \limits^{(c)} &{\rm{  }}{\boldsymbol{\mu} ^H}{{\bf{h}}_k}{{\bf{w}}_k},
\end{split}
\end{equation}
where $(a)$ holds by defining ${\boldsymbol{\theta}_r} = {\left[{{\theta _{r,1}} ,\ldots, {\theta _{r,N}}} \right]^{{\rm{ }}T}}$, $(b)$ holds by defining $\boldsymbol{\theta} = {\left[{\boldsymbol{\theta} _1^T ,\ldots, \boldsymbol{\theta} _R^T}\right]^{{\rm{ }}T}}$, and ${{\bf{F}}_k} = {\left[ {{\bf{F}}_{1,k}^T ,\ldots, {\bf{F}}_{R,k}^T} \right]^{{\rm{ }}T}}$, ${{\bf{G}}_b} = {\left[ {{\bf{G}}_{b,1}^T ,\ldots, {\bf{G}}_{b,R}^T} \right]^{{\rm{ }}T}}$, and $(c)$ holds by defining $\boldsymbol{\mu}  = {\left[ {{\boldsymbol{\theta} ^T},1} \right]^{{\rm{ }}T}}$, ${{\bf{w}}_k} = {\left[ {{\bf{w}}_{1,k}^T ,\ldots, {\bf{w}}_{b,k}^T} \right]^{{\rm{ }}T}}$, and
${{\bf{h}}_k} = \left[{\begin{array}{c}
{{\rm{diag}}({\bf{F}}_k^H){{\bf{G}}_1}} \cdots {{\rm{diag}}({\bf{F}}_k^H){{\bf{G}}_b}}\\
{{\bf{H}}_{1,k}^H}\ \ \ \ \cdots \ \ \ \ {{\bf{H}}_{b,k}^H}
\end{array}} \right]$.
Similarly, we have
\begin{equation}\label{eq:6}
\begin{split}
&{\rm{     }}\sum\limits_{b = 1}^B \left( {\bf{H}}_{b,e}^H + \sum\limits_{r = 1}^R {{\bf{F}}_{r,e,k}^H} {\bf{\Theta }}_r^H{{\bf{G}}_{b,r}}\right){{\bf{w}}_{b,k}}\\
\mathop  = \limits^{(d)} &{\rm{ }}\sum\limits_{b = 1}^B \left( {\bf{H}}_{b,e}^H + {\boldsymbol{\theta} ^H}{\rm{diag}}\left({\bf{F}}_{e,k}^H\right){{\bf{G}}_{b}}\right){{\bf{w}}_{b,k}}\\
\mathop  = \limits^{(e)} &{\rm{  }}{\boldsymbol{\mu}^H}{{\bf{h}}_{e,k}}{{\bf{w}}_k},
\end{split}
\end{equation}
where $(d)$ holds by defining ${{\bf{F}}_{e,k}} = {\left[ {{\bf{F}}_{1,e,k}^T ,\ldots, {\bf{F}}_{R,e,k}^T} \right]^{{\rm{ }}T}}$, and
${{\bf{h}}_{e,k}} = \left[{\begin{array}{c}
{{\rm{diag}}({\bf{F}}_{e,k}^H){{\bf{G}}_1}} \cdots {{\rm{diag}}({\bf{F}}_{e,k}^H){{\bf{G}}_b}}\\
{{\bf{H}}_{1,e}^H}\ \ \ \ \cdots \ \ \ \ {{\bf{H}}_{b,e}^H}
\end{array}} \right]$.
Then, the signal-to-interference-plus-noise ratio (SINR) of the received signal of the $k$-th user and the Eve eavesdropping the $k$-th user's information can be calculated as
\begin{subequations}\label{eq:p7}
\begin{align}
{\gamma _k} &= \frac{{{{\left| {{\boldsymbol{\mu} ^H}{{\tilde{\mathbf{h}}}_k}{{\bf{w}}_k}} \right|}^2}}}{{\sum\limits_{j = 1,j \ne k}^K {{{\left| {{\boldsymbol{\mu} ^H}{{\tilde{\mathbf{h}}}_k}{{\bf{w}}_j}} \right|}^2} + 1} }},\\
{\gamma _k^e} &= \frac{{{{\left| {{\boldsymbol{\mu} ^H}{{\tilde{\mathbf{h}}}_{e,k}}{{\bf{w}}_k}} \right|}^2}}}{{\sum\limits_{j = 1,j \ne k}^K {{{\left| {{\boldsymbol{\mu} ^H}{{\tilde{\mathbf{h}}}_{e,k}}{{\bf{w}}_j}} \right|}^2} + 1} }},
\end{align}
\end{subequations}
where ${\tilde{\mathbf{h}} _k}={\sigma _k^{-1}}{{\mathbf{h}} _k}$ and ${\tilde{\mathbf{h}} _{e,k}}={\sigma _e^{-1}}{{\mathbf{h}} _{e,k}}$.

\subsection{Problem Formulation}
Thus, the achievable secrecy rate from BSs to the $k$-th user can be expressed as
\begin{equation}\label{eq:p8}
{R_{s,k}} = {\left[{\kern 1pt} {\ln (1 + {\gamma _k}) - \ln (1 + {\gamma _{e,k}})}{\kern 1pt} \right]^ + },
\end{equation}
where ${\left[ z \right]^ + } = \max \left({z,0}\right)$.

In this paper, we aim to maximize the WSSR by jointly optimizing the BF vector ${\bf{w}}_k$ and the phase shift $\boldsymbol{\mu}$. Mathematically, the optimization problem can be formulated as follows

\begin{subequations}\label{eq:p9}
\begin{align}
{{\cal P}^o}:\mathop {\max }\limits_{{\bf{W}},\boldsymbol{\mu}} {\rm{  }} \ &{R_s} = \sum\limits_{k = 1}^K {{\eta _k}(\ln {\kern 1pt} (1 + {\gamma _k}) - \ln {\kern 1pt}  (1 + \gamma _k^e)} )\\
{\rm{s}}{\rm{.t.}} \ &C1:{\sum\limits_{k = 1}^K} {{\rm{ }}{{\left\| {{{\bf{w}}_{b,k}}} \right\|}^{{\kern 1pt} 2}}}  \le {P_b}{\rm{ }},\forall b \in {\cal B},\\
&C2:{\rm{ }}\left| {{\theta _{r,n}}} \right| = 1{\rm{ }},{\rm{ }}\forall r \in {\cal R},\forall n \in {\cal N},
\end{align}
\end{subequations}
where ${\eta _k}\left(0 \le {\eta _k} \le 1 \right)$ represents the weight of the $k$-th user, ${P_b}$ represents the maximum available transmit power of the $b$-th BS, and ${\bf{W}}$ is defined as ${\bf{W}}{\rm{ = }}{\left[ {{\bf{w}}_1^T, \ldots ,{\bf{w}}_K^T} \right]^T}$. Note that in (9a) we omit ${\left[  \cdot  \right]^ + }$, because the optimal value should be non-negative.

Due to the coupling of ${\bf{W}}$ and $\boldsymbol{\mu}$ in the objective function (9a) and the unit modulus constraint of $C2$, it is difficult to solve problem ${\cal P}^o$. Here we take the AO method to decouple variables
${\bf{W}}$ and $\boldsymbol{\mu}$, and decompose the original problem ${\cal P}^o$ into two sub-problems. We then fix other variables in each iteration of the sub-problems and solve each convex problem using SDR and SCA methods. The detail will be given in the following section. 

\section{Joint Optimization Framework}
In this section, we will systematically describe the AO algorithm for solving the problem ${\cal P}^o$, namely by iteratively solving two sub-problems (one is to optimize ${\bf{W}}$ with a fixed $\boldsymbol{\mu}$, while the other is to optimize $\boldsymbol{\mu}$ with a fixed ${\bf{W}}$), as shown in \textbf{Algorithm 1}.
\subsection{Sub-Problem for Solving BF ${\bf{W}}$}
Firstly, we fix the phase shift variable $\boldsymbol{\mu}$ and solve the BF matrix ${\bf{W}}$ in the original problem ${\cal P}^o$. Here, we give
the following two inequalities near the fixed point $\left\{ {\hat a,\hat b} \right\}$:
\begin{subequations}\label{eq:p10}
\begin{align}
\ln \left( {1 + \frac{{{{\left| a \right|}^2}}}{b}} \right) \ge \ln \left( {1 + \frac{{{{\left| {\hat a} \right|}^2}}}{{\hat b}}} \right) - \frac{{{{\left| {\hat a} \right|}^2}}}{{\hat b}} \ \ \ \ \ \ \ \ \nonumber \\
 + \frac{{2\Re \left\{ {a\hat a} \right\}}}{{\hat b}} - \frac{{{{\left| {\hat a} \right|}^2}\left( {b + {{\left| a \right|}^2}} \right)}}{{\hat b{\rm{ }}\left( {\hat b + {{\left| {\hat a} \right|}^2}} \right)}}, \\
\ln \left( {1 + \frac{a}{b}} \right) \le  \left( {1 + \frac{{\hat a}}{{\hat b}}} \right) - \frac{{\hat b}}{{\hat b + \hat a}}{\kern 1pt} \left( {\frac{a}{b} - \frac{{\hat a}}{{\hat b}} } \right).
\end{align}
\end{subequations}

\begin{algorithm}
	\renewcommand{\algorithmicrequire}{\textbf{Input:}}
	\renewcommand{\algorithmicensure}{\textbf{Output:}}
    \caption{Proposed Joint Optimization AO Algorithm.}
    \label{Algorithm1}
    \begin{algorithmic}[1]
        \REQUIRE ${{\bf{H}}_{b,k}}$, ${{\bf{H}}_{b,e}}$, ${{\bf{F}}_{r,k}}$, ${{\bf{F}}_{r,e,k}}$, ${{\bf{G}}_{b,r}}$, and ${\sigma _k}$, ${\sigma _e}$.
        \ENSURE BF matrix ${\bf{W}}$, phase shift vector $\boldsymbol{\mu}$, and the maximum WSSR ${R_s}$.
        \STATE Initialize ${\bf{W}}^t$, ${\boldsymbol{\mu}}^t$, ${R_s^t}$, $t=0$ and threshold $\varepsilon$;
        \WHILE{ ${R_s^t}-{R_{s}^{t-1}}>{\varepsilon}$}
        \STATE $t=t+1$;
        \STATE Update ${\bf{W}}^t$ by solving ${{\cal P}^1}$;
        \STATE Update $\boldsymbol{\mu}^t$ by solving ${{\cal P}^5}$;
        \STATE Update ${R_s^t}$ by (9a);
        \ENDWHILE
        \STATE \textbf{return} ${\bf{W}}^t$, ${\boldsymbol{\mu}}^t$, and ${R_s^t}$.
    \end{algorithmic}
\end{algorithm}

Based on the above, we have the following inferences. Firstly, the information rate of the $k$-th user with fixed point $\left\{ {{\bf{w}}_k^t,{\boldsymbol{\mu} ^t}} \right\}$ can be approximated as
\begin{equation}\label{eq:11}
\begin{split}
\ln \left( {1 + {\gamma _k}} \right) =& \ln \left( {1 + \frac{{{{\left| {{{\left( {{\boldsymbol{\mu} ^t}} \right)}^H}{{\tilde{\mathbf{h}}}_k}{{\bf{w}}_k}} \right|}^2}}}{{\sum\limits_{j = 1,j \ne k}^K {{{\left| {{{\left( {{\boldsymbol{\mu} ^t}} \right)}^H}{{\tilde{\mathbf{h}}}_k}{{\bf{w}}_j}} \right|}^2} + 1} }}} \right) \\
 \ge &\ln \left( {1 + \frac{{{{\left| {\alpha _k^t} \right|}^2}}}{{\beta _k^t}}} \right)  + \frac{{2\Re \left\{ {\alpha _k^t{\alpha _k}} \right\}}}{{\beta _k^t}} \\
 &- \frac{{{{\left| {\alpha _k^t} \right|}^2}}}{{\beta _k^t}}- \frac{{{{\left| {\alpha _k^t} \right|}^2}\left( {{\beta _k} + {{\left| {{\alpha _k}} \right|}^2}} \right)}}{{\beta _k^t{\rm{ }}\left( {\beta _k^t + {{\left| {\alpha _k^t} \right|}^2}} \right)}},
\end{split}
\end{equation}
where
\begin{subequations}\label{eq:p12}
\begin{align}
& {\alpha _k} = {\left( {{\boldsymbol{\mu} ^t}} \right)^H}{{\tilde{\mathbf{h}}}_k}{{\bf{w}}_k},\qquad\\
& \alpha _k^t = {\left( {{\boldsymbol{\mu} ^t}} \right)^H}{{\tilde{\mathbf{h}}}_k}{\bf{w}}_k^t,\quad\\
& {\beta _k} = \sum\limits_{j = 1,j \ne k}^K {{{\left| {{{\left( {{\boldsymbol{\mu} ^t}} \right)}^H}{{\tilde{\mathbf{h}}}_k}{{\bf{w}}_j}} \right|}^2} + 1},\quad \\
& \beta _k^t = \sum\limits_{j = 1,j \ne k}^K {{{\left| {{{\left( {{\boldsymbol{\mu} ^t}} \right)}^H}{{\tilde{\mathbf{h}}}_k}{\bf{w}}_j^t} \right|}^2} + 1}.\quad
\end{align}
\end{subequations}

Then, the rate of negative weight of the Eve eavesdropping the $k$-th user's information can be written as
\begin{equation}\label{eq:13}
\begin{split}
- \ln \left( {1 + \gamma _k^e} \right) =& \ln \left( {1 + \sum\limits_{j = 1,j \ne k}^K {{{\left| {{{\left( {{\boldsymbol{\mu} ^t}} \right)}^H}{{\tilde{\mathbf{h}}}_{e,k}}{{\bf{w}}_j}} \right|}^2}} } \right) \\
&- \ln \left( {1 + \sum\limits_{j = 1}^K {{{\left| {{{\left( {{\boldsymbol{\mu} ^t}} \right)}^H}{{\tilde{\mathbf{h}}}_{e,k}}{{\bf{w}}_j}} \right|}^2}} } \right).
\end{split}
\end{equation}

The first term of the (13) can be approximated as
\begin{equation}\label{eq:14}
\begin{split}
 &\ln \left( {1 + {{\bf{W}}^H}{{\bf{\Omega }}_k}{\bf{\Omega }}_k^H{\bf{W}}} \right)    \\  \vspace{2ex}
 &\ge {\rm{ln}}\left( {1 + \left\| {{\bf{\Omega }}_k^H{{\bf{W}}^t}} \right\|{^2}} \right) + 2\Re \left\{ {{{\bf{W}}^H}{{\bf{\Omega }}_k}{\bf{\Omega }}_k^H{{\bf{W}}^t}} \right\}  \\
 &\ \ \ \  - \left\| {{\bf{\Omega }}_k^H{{\bf{W}}^t}} \right\|{^2} - \frac{{\left\| {{\bf{\Omega }}_k^H{{\bf{W}}^t}} \right\|^2}\left( {1 + {{\left\| {{\bf{\Omega }}_k^H{\bf{W}}} \right\|}^2}} \right)}{{{\rm{ }}1 + \left\| {{\bf{\Omega }}_k^H{{\bf{W}}^t}} \right\|^2}},
\end{split}
\end{equation}
where
\begin{equation}\label{eq:15}
{{\bf{\Omega }}_k}{\bf{\Omega }}_k^H = {\rm{diag}}\left[ {{\boldsymbol{\xi} _{e,k}}, \ldots ,\underbrace {{{\bf{0}}_{MB \times MB}}}_{k - {\rm{th \ term}}}, \ldots ,{\kern 1pt} {\boldsymbol{\xi}_{e,k}}} \right].
\end{equation}
In (15), ${\boldsymbol{\xi} _{e,k}}$ is calculated as ${\boldsymbol{\xi} _{e,k}} = {\tilde{\mathbf{h}}}_{e,k}^H{\boldsymbol{\mu} ^t}{\left( {{\boldsymbol{\mu} ^t}} \right)^H}{{\tilde{\mathbf{h}}}_{e,k}}$. The second term on the right side of (13) can be approximately as
\begin{equation}\label{eq:16}
 - \ln \left( {1 + {\chi _k}} \right) \ge  - \ln \left( {1 + \chi _k^t} \right) - \frac{{1 + {\chi _k}}}{{1 + \chi _k^t}} + 1
 \end{equation}
 where
 \begin{subequations}\label{eq:p17}
\begin{align}
&{\chi _k} = \sum\limits_{j = 1}^K {{{\left| {{{\left( {{\boldsymbol{\mu} ^t}} \right)}^H}{{\tilde{\mathbf{h}}}_{e,k}}{{\bf{w}}_j}} \right|}^2}} {\rm{ }},\\
&\chi _k^t = \sum\limits_{j = 1}^K {{{\left| {{{\left( {{\boldsymbol{\mu} ^t}} \right)}^H}{{\tilde{\mathbf{h}}}_{e,k}}{\bf{w}}_j^t} \right|}^2}}.
\end{align}
\end{subequations}

Finally, based on the above inferences, the WSSR maximization problem ${{\cal P}^o}$  can be reformulated as the following sub-problem ${{\cal P}^1}$:
\begin{subequations}\label{eq:p18}
\begin{align}
{{\cal P}^1}: &\mathop {\min }\limits_{\bf{W}}  \ \sum\limits_{k = 1}^K {{\eta _k}}  \left\{{ - \frac{{2\Re \left\{ {\alpha _k^t{\alpha _k}} \right\}}}{{\beta _k^t}} + \frac{{{{\left| {\alpha _k^t} \right|}^2}\left( {{\beta _k} + {{\left| {{\alpha _k}} \right|}^2}} \right)}}{{\beta _k^t{\rm{ }}\left( {\beta _k^t + {{\left| {\alpha _k^t} \right|}^2}} \right)}}} \right. \nonumber  \\
&\qquad \qquad \qquad  + \frac{{{\chi _k}}}{{\chi _k^t + 1}} - 2\Re \left\{ {{{\bf{W}}^H}{{\bf{\Omega }}_k}{\bf{\Omega }}_k^H{{\bf{W}}^t}} \right\}   \nonumber \\
&\qquad \qquad \qquad   \left.+ \frac{{{\kern 1pt} {\kern 1pt} \left\| {{\bf{\Omega }}_k^H{{\bf{W}}^t}} \right\|^2 {{{\left\| {{\bf{\Omega }}_k^H{\bf{W}}} \right\|}^2}} {\kern 1pt} }}{{{\rm{ }}1 + \left\| {{\bf{\Omega }}_k^H{{\bf{W}}^t}} \right\|{\kern 1pt} {{\kern 1pt} ^2}}} \right\}  \\
&{\rm{s}}{\rm{.t}}{\rm{.   }} \ C1 \ {\rm{: }} \ \sum\limits_{k = 1}^K {{\rm{ }}{{\left\| {{{\bf{w}}_{b,k}}} \right\|}^{{\kern 1pt} 2}}}  \le {P_b}{\rm{ , }}\forall b \in {\cal B}.\qquad
\end{align}
\end{subequations}
Here, ${{\cal P}^1}$ is a convex problem, which can be solved by standard convex optimization techniques or CVX toolbox.
\subsection{Sub-Problem for Solving Phase Shift ${\boldsymbol{\mu}}$}
Similarly, by fixing the BF matrix ${\bf{W}}$, we can rewrite problem ${{\cal P}^o}$ as a sub-problem w.r.t. ${\boldsymbol{\mu}}$ around the fixed point $\left\{ {{\bf{w}}_k^t,{\boldsymbol{\mu} ^t}} \right\}$  as follows:
\begin{subequations}\label{eq:p19}
\begin{align}
{{\cal P}^2}:&\mathop {\min }\limits_{\boldsymbol{\mu}} {\rm{  }} \  \sum\limits_{k = 1}^K {\eta _k}   \left\{ - \frac{{2\Re \left\{ {\alpha _k^t{\alpha _k}} \right\}}}{{\beta _k^t}} + \frac{{{{\left| {\alpha _k^t} \right|}^2}\left( {{\beta _k} + {{\left| {{\alpha _k}} \right|}^2}} \right)}}{{\beta _k^t{\rm{ }}\left( {\beta _k^t + {{\left| {\alpha _k^t} \right|}^2}} \right)}} \right.   \nonumber \\
&+ \left. \frac{{{\chi _k}}}{{\chi _k^t + 1}} - \ln \left( {1 + \sum\limits_{j = 1,j \ne k}^K {{{\left| {{{\left( {{\boldsymbol{\mu} ^t}} \right)}^H}{{\tilde{\mathbf{h}}}_{e,k}}{{{\bf{w}}_j^t}}} \right|}^2}} } \right) \right\}\\
&{\rm{s}}{\rm{.t.}} \ C3:\left| {{\boldsymbol{\mu} _n}} \right| = 1,\forall n \in \left\{ {1, \ldots ,NR} \right\} ,   \nonumber \\
&\qquad \qquad  {\boldsymbol{\mu} _{NR + 1}} = 1,
\end{align}
\end{subequations}
where the calculation of ${\alpha _k^t}$, ${\beta _k^t}$ and $\chi _k^t$ is the same as (12b), (12d) and (17b) in the previous sub-section. Besides,
\begin{subequations}\label{eq:p20}
\begin{align}
&{\alpha _k} = {\boldsymbol{\mu} ^H}{{\tilde{\mathbf{h}}}_k}{\bf{w}}_k^t,\\
&{\beta _k} = \sum\limits_{j = 1,j \ne k}^K {{{\left| {{\boldsymbol{\mu} ^H}{{\tilde{\mathbf{h}}}_k}{{\bf{w}}_j^t}} \right|}^2} + 1},\\
&{\chi _k} = \sum\limits_{j = 1}^K {{\kern 1pt} {{\left| {{\boldsymbol{\mu} ^H}{{\tilde{\mathbf{h}}}_{e,k}}{{\bf{w}}_j^t}} \right|}^2}} .
\end{align}
\end{subequations}

For $\ln \left( {1 + \sum\limits_{j = 1,j \ne k}^K {{{\left| {{{\left( {{\boldsymbol{\mu} ^t}} \right)}^H}{{\tilde{\mathbf{h}}}_{e,k}}{{{\bf{w}}_j^t}}} \right|}^2}} } \right)$ in (19a), it can also be approximated as
\begin{equation}\label{eq:21}
\begin{split}
&\ln  \left( {1 + {\boldsymbol{\mu} ^H}{{\bf{\Psi }}_k}{\bf{\Psi }}_k^H\boldsymbol{\mu} } \right)  \\
 &\ge {\rm{ln}}\left( {1 + {{\left\| {{\bf{\Psi }}_k^H{\boldsymbol{\mu} ^t}} \right\|}^2}} \right) + 2\Re \left\{ {{\boldsymbol{\mu} ^H}{{\bf{\Psi }}_k}{\bf{\Psi }}_k^H{\boldsymbol{\mu} ^t}} \right\}  \\
  &\ \ \ - {\left\| {{\bf{\Psi }}_k^H{\boldsymbol{\mu} ^t}} \right\|^2}  - \frac{{{\kern 1pt} {\kern 1pt} {{\left\| {{\bf{\Psi }}_k^H{\boldsymbol{\mu} ^t}} \right\|}^2}{\kern 1pt} \left( {1 + {{\left\| {{\bf{\Psi }}_k^H\boldsymbol{\mu} } \right\|}^2}} \right)}}{{{\rm{ }}1 + \left\| {{\bf{\Psi }}_k^H{\boldsymbol{\mu} ^t}} \right\|{\kern 1pt} {{\kern 1pt} ^2}}},
\end{split}
\end{equation}
where ${{\bf{\Psi }}_k}{\bf{\Psi }}_k^H = \sum\limits_{j = 1,j \ne k}^K {{{\tilde{\mathbf{h}}}_{e,k}}{\bf{w}}_j^t{{({\bf{w}}_j^t)}^H}{\tilde{\mathbf{h}}}_{e,k}^H}$. Omitting the constant term, we can rewrite the problem ${{\cal P}^2}$ as
\begin{subequations}\label{eq:p22}
\begin{align}
{{\cal P}^3}:&\mathop {\min }\limits_{\boldsymbol{\mu}} {\rm{  }} \  \sum\limits_{k = 1}^K {\eta _k}   \left\{ - \frac{{2\Re \left\{ {\alpha _k^t{\alpha _k}} \right\}}}{{\beta _k^t}} + \frac{{{{\left| {\alpha _k^t} \right|}^2}\left( {{\beta _k} + {{\left| {{\alpha _k}} \right|}^2}} \right)}}{{\beta _k^t{\rm{ }}\left( {\beta _k^t + {{\left| {\alpha _k^t} \right|}^2}} \right)}} \right.   \nonumber \\
&\qquad \qquad \qquad+ \frac{{{\chi _k}}}{{\chi _k^t + 1}} -  2\Re \left\{ {{\boldsymbol{\mu} ^H}{{\bf{\Psi }}_k}{\bf{\Psi }}_k^H{\boldsymbol{\mu} ^t}} \right\}  \nonumber\\
&\qquad \qquad \qquad + \left.  \frac{{{\kern 1pt} {\kern 1pt} {{\left\| {{\bf{\Psi }}_k^H{\boldsymbol{\mu} ^t}} \right\|}^2}{\kern 1pt} { {{\left\| {{\bf{\Psi }}_k^H \boldsymbol{\mu} } \right\|}^2}} }}{{{\rm{ }}1 + \left\| {{\bf{\Psi }}_k^H{\boldsymbol{\mu} ^t}} \right\|{\kern 1pt} {{\kern 1pt} ^2}}} \right\} \\
&{\rm{s}}{\rm{.t.}} \ C3.
\end{align}
\end{subequations}
For brevity, we denote $\boldsymbol{\varpi} = \sum\limits_{j = 1}^K {{\bf{w}}_j^t{{({\bf{w}}_j^t)}^H}} $, and then ${{\beta _k} + {{\left| {{\alpha _k}} \right|}^2}}={\boldsymbol{\mu} ^H}{{\tilde{\mathbf{h}}}_k}\boldsymbol{\varpi} {\tilde{\mathbf{h}}}_k^H\boldsymbol{\mu}  + 1$, ${\chi _k} = {\boldsymbol{\mu} ^H}{{\tilde{\mathbf{h}}}_{e,k}}\boldsymbol{\varpi} {\tilde{\mathbf{h}}}_{e,k}^H\boldsymbol{\mu} $.
Next, we denote
\begin{subequations}\label{eq:p23}
\begin{align}
&{\mathbf{A}_k}{\rm{ = }}\frac{{{{\left| {\alpha _k^t} \right|}^2}\left( {{{\tilde{\mathbf{h}}}_k}\boldsymbol{\varpi} {\tilde{\mathbf{h}}}_k^H} \right)}}{{\beta _k^t{\rm{ }}\left( {\beta _k^t + {{\left| {\alpha _k^t} \right|}^2}} \right)}} + \frac{{{{\tilde{\mathbf{h}}}_{e,k}}\boldsymbol{\varpi} {\tilde{\mathbf{h}}}_{e,k}^H}}{{\chi _k^t + 1}} + \frac{{{\kern 1pt} {\kern 1pt} {{\bf{\Psi }}_k}{\bf{\Psi }}_k^H{{\left\| {{\bf{\Psi }}_k^H{\boldsymbol{\mu} ^t}} \right\|}^2}}}{{{\rm{ 1 + }}{\kern 1pt} {\kern 1pt} {{\left\| {{\bf{\Psi }}_k^H{\boldsymbol{\mu} ^t}} \right\|}^2}}},\\
& \qquad \qquad \qquad {\mathbf{v}_k}{\rm{ = }}{\kern 1pt}{\kern 1pt}{{\bf{\Psi }}_k}{\bf{\Psi }}_k^H{\boldsymbol{\mu} ^t} + \frac{{{{\tilde{\mathbf{h}}}_k}{\bf{w}}_k^t\alpha _k^t}}{{\beta _k^t}}.
\end{align}
\end{subequations}
Hence, we can simplify problem ${{\cal P}^3}$ as follows:
\begin{subequations}\label{eq:p24}
\begin{align}
{{\cal P}^4}:&\mathop {\min }\limits_{\boldsymbol{\mu}} \  {\rm{   }}{\boldsymbol{\mu} ^H}{\bf{A}}{\kern 1pt} \boldsymbol{\mu}  - 2\Re \left\{ {{\boldsymbol{\mu} ^H}{\bf{v}}} \right\}\\
&{\kern 1pt} {\kern 1pt} {\rm{s}}{\rm{.t.}} \ C3,
\end{align}
\end{subequations}
where ${\bf{A}} = \sum\limits_{k = 1}^K {{\eta _k}{{\bf{A}}_k}}$, and ${\bf{v}} = \sum\limits_{k = 1}^K {{\eta _k}} {{\bf{v}}_k}$.

The simplified sub-problem ${{\cal P}^4}$ is a quadratic programming (QP) problem with unit modulus constraint. Next, we will employ the alternating direction method of multipliers (ADMM) technology to solve it. Firstly, we introduce a slack variable $\mathbf{p} \in {{\mathbb{C}}^{NR+1 }}$, and the problem ${{\cal P}^4}$ can be rewritten as
\begin{subequations}\label{eq:p25}
\begin{align}
{{\cal P}^5}:&\mathop {\min }\limits_{{ \mathbf{p}},\boldsymbol{\mu} } \ {\rm{   }}{{\mathbf{p}}^H}{\bf{A}}{\kern 1pt} {\mathbf{p}} - 2\Re \{ {{\mathbf{p}}^H}{\bf{v}}\} \\
&{\rm{s}}{\rm{.t.}} \ \ {\mathbf{p}}= \boldsymbol{\mu} ,\\
 &\quad \ \ {\kern 1pt}{\kern 1pt} C3.
\end{align}
\end{subequations}
The augmented Lagrangian equation of problem ${{\cal P}^5}$ can be expressed as
\begin{equation}\label{eq:26}
\begin{split}
\mathbf{{\cal G}}\left( {{\mathbf{p}},\boldsymbol{\mu} ,\boldsymbol{\lambda} } \right) =& {{\mathbf{p}}^H}{\bf{A}}{\kern 1pt} {\mathbf{p}} - 2\Re \{ {{\mathbf{p}}^H}{\bf{v}}\}  \\
&- \Re \left\{ {\boldsymbol{\lambda} \left( {{\mathbf{p}} - \boldsymbol{\mu}} \right)} \right\} + \frac{\delta }{2}{\left\| {{\mathbf{p}} - \boldsymbol{\mu} } \right\|^2} ,
\end{split}
\end{equation}
where $\boldsymbol {\lambda} \in {{\mathbb{C}}^{NR+1 }}$ and ${\delta }\geq 0$ represent Lagrangian multipliers. We put the superscript $t$ of each variable to represent the index of the iteration, and then the process of each iteration is as follows:
\begin{subequations}\label{eq:p27}
\begin{align}
&{{\mathbf{p}}^{t + 1}} = \arg {\kern 1pt} {\kern 1pt} \mathop {\min }\limits_{\mathbf{p}} {\kern 1pt} {\kern 1pt} {\kern 1pt} {\kern 1pt} {\kern 1pt} {\mathbf{{\cal G}}}\left( {{{\mathbf{p}}^t},{\boldsymbol{\mu}^t},{\boldsymbol {\lambda} ^t}} \right), \\
&{\boldsymbol{\mu} ^{t + 1}} = \arg {\kern 1pt} {\kern 1pt} \mathop {\min }\limits_{\boldsymbol{\mu}}  {\kern 1pt} {\kern 1pt} {\kern 1pt} {\kern 1pt} {\kern 1pt} {\mathbf{{\cal G}}}\left( {{{\mathbf{p}}^{t + 1}},{\boldsymbol{\mu} ^t},{\boldsymbol {\lambda} ^t}} \right)  ,\\
&{\boldsymbol {\lambda} ^{t + 1}} = {\boldsymbol {\lambda} ^t} - \delta \left( {{{\mathbf{p}}^{t + 1}} - {\boldsymbol{\mu} ^{t + 1}}} \right) .
\end{align}
\end{subequations}

\emph{1) Optimizing} ${\mathbf{p}}$: Here, the first-order optimization solves the gradient $\partial {\mathbf{{\cal G}}}/\partial {\boldsymbol p}$ as follows:
\begin{equation}\label{eq:28}
\frac{{\partial  {\mathbf{{\cal G}}}}}{{\partial {\mathbf{p}}}} = 2{\bf{A}}{{\mathbf{p}}^{t + 1}} - 2{\bf{v}} - {\boldsymbol {\lambda} ^t} + \delta {\rm{(}}{{\mathbf{p}}^{t + 1}} - {\boldsymbol{\mu} ^t}{\rm{)}}.
\end{equation}
Let (28) be equal to zero, we can obtain
\begin{equation}\label{eq:29}
 {{\mathbf{p}}^{t + 1}} = (2{\bf{A}} + \delta {\bf{I}}{\rm{)}}{{\kern 1pt} ^{ - 1}}{\kern 1pt} (2{\bf{v}} + {\boldsymbol {\lambda} ^t} + \delta {\boldsymbol{\mu} ^t}).
\end{equation}

\emph{2) Optimizing} ${\boldsymbol {\mu}}$: The solution of (27b) is similar to the above, but with the constraint of (25c). Its solution is equivalent to
\begin{equation}\label{eq:30}
\mathop {\min }\limits_{\left| {{\boldsymbol{\mu} _n}} \right| = 1,{\boldsymbol{\mu} _{NR + 1}} = 1} {\kern 1pt} {\kern 1pt} {\kern 1pt} {\kern 1pt} {\kern 1pt} \left\| {\boldsymbol{\mu} - ({{\mathbf{p}}^{t + 1}} - {\delta ^{ - 1}}{\boldsymbol {\lambda} ^t})} \right\| ,
\end{equation}
i.e.,
\begin{equation}\label{eq:31}
\begin{split}
{\left[ {{\boldsymbol{\mu} ^{t + 1}}} \right]_{{\kern 1pt} n}} = \left\{ \begin{array}{l}
 \frac{{{{\left[ {{\kern 1pt} {{\mathbf{p}}^{t + 1}} - {\delta ^{ - 1}}{\boldsymbol {\lambda} ^t}} \right]}_{{\kern 1pt} n}}}}{{\left| {{{\left[ {{\kern 1pt} {{\mathbf{p}}^{t + 1}} - {\delta ^{ - 1}}{\boldsymbol {\lambda} ^t}} \right]}_{ n}}} \right|}} ,\ \ {\left[ {{\kern 1pt} {{\mathbf{p}}^{t + 1}} - {\delta ^{ - 1}}{\boldsymbol {\lambda} ^t}} \right]_{{\kern 1pt} n}} \ne 0,\\
\\
 {\left[ {{\boldsymbol{\mu} ^t}} \right]_{ n}} , \qquad \qquad \quad {\rm{           }}{\left[ {{\kern 1pt} {{\mathbf{p}}^{t + 1}} - {\delta ^{ - 1}}{\boldsymbol {\lambda} ^t}} \right]_{{\kern 1pt} n}} = 0,{\rm{ }}
\end{array} \right.
\end{split}
\end{equation}
where $\forall n \in \left\{ {1, \ldots ,NR} \right\}$, and ${\left| {{\boldsymbol{\mu} ^{t + 1}}} \right|_{NR + 1}} = 1$.

\emph{3) Optimizing} ${\boldsymbol {\lambda}}$: From (28), we can solve (27c) as
\begin{equation}\label{eq:32}
{\boldsymbol {\lambda} ^{t + 1}} = 2{\bf{A}}{{\mathbf{p}}^{t + 1}} - 2{\bf{v}}.
\end{equation}

The above three sub-problems can be solved in closed form. Finally, to make the ADMM algorithm converge, we have the following lemma [30].

\textbf{Lemma 1}: The above ADMM algorithm converges, no matter whether $\boldsymbol{\mu}$ belongs to a continuous set, as long as the penalty factor $\delta$ satisfies:
\begin{equation}\label{eq:33}
\frac{\delta }{2}{\bf{I}} - {\bf{A}} \succ {\bf{0}}.
\end{equation}
\emph{Proof}: The certification process can refer to [32].

Here, we briefly summarize the proposed scheme for solving the original problem ${\cal P}^o$. First, we solve sub-problem ${\cal P}^1$ by fixing $\boldsymbol{\mu}$, and then solve sub-problem ${\cal P}^5$ based on obtaining $\mathbf{W}$ from the ${\cal P}^1$. For Solving ${\cal P}^5$, the ADMM method is applied to alternately iterating (27a), (27b), and (27c) until the result converges. Next, ${\cal P}^1$ is re-solved based on obtaining $\boldsymbol{\mu}$ from the ${\cal P}^5$, and the above process is repeated until convergence.

 \begin{figure}[t]
  \centering
  \includegraphics[scale = 0.36]{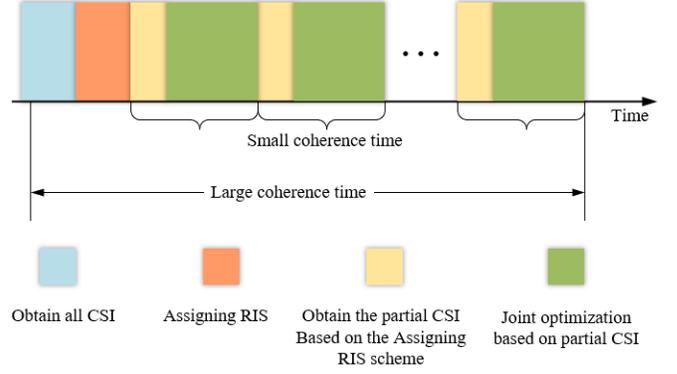}
  \caption{The work process of the joint optimization of assigning RIS extension framework.}
  \label{fig.2}
\end{figure}

\section{Extension of the Joint Optimization Framework for Assigning RIS}
In previous section, we have proposed the joint optimization framework based on perfect CSI. However, due to the intrinsic high-dimensional form of the RIS channel and many RISs, BSs and users in RIS-aided cell-free network, it is impractical or extremely difficult to obtain CSI of RISs so frequently.

Therefore, here we propose an extended scheme for assigning RISs to users, and then reconsider the WSSR maximization problem. In order to facilitate the description, the concept of two coherent times is introduced as shown in Fig. 2, which is similar to [25]. The proposed \textbf{Algorithm 1} is based on the obtained CSI at each coherence time. However, the
proposed assigning RIS scheme is to obtain all CSI at the start of the large coherence time, and select a few RISs that can provide good secure service quality for the user.
After that, we only need the CSI of selected RISs in each subsequent small coherence time, and perform joint optimization problem based on the selected RISs.

\subsection{Overview of Assigning RIS}
In this section, we propose an assigning RIS optimization scheme that balances the security performance and cost as shown in Fig. 2. Specifically, we select RISs with better secure service quality for joint optimization based on the feature that the RIS with poor service quality has less impact on user secrecy rate. At the start of a large coherence time, a full channel estimation is performed and the RIS with better service quality is selected for each user. After that, in each repeated small coherence time, we only obtain the CSI related to the allocated RIS and perform joint optimization problem. Here we omit the channel estimation of RISs with poor service quality in each small coherence time, which greatly decreases the overhead with a small performance loss. Meanwhile, those RISs that are not screened for user $k$ will not provide services to it. Because the phase shift of the unscreened RIS without joint optimization is actually equivalent to a random phase shift, the impact on the user's secrecy rate is very limited. The specific performance loss and impact will be shown in detail later in Section VI.
\subsection{Problem Formulation of Assigning RIS}
To determine the matching problem among RISs and users, we introduce a matching factor variable $l_{r,k}$ $\left(l\in\{0,1\}\right)$ to indicate whether the $k$-th user and the $r$-th RIS are grouped. When $l_{r,k}$ is $0$, it means that the $r$-th RIS does not serve the $k$-th user and vice versa. For convenience, we define $\textbf{L}_k = [{l_{1,k}}, \ldots ,{l_{R,k}}]^T$. To express the matching problem, we define the virtual received signal of the $k$-th user and the Eve eavesdropping on the $k$-th user as follows
\begin{subequations}\label{eq:p34}
\begin{align}
{{{{\hat y}}}_k} = \sum\limits_{b = 1}^B {\sum\limits_{j = 1}^K {\left( {{\bf{H}}_{b,k}^H + \sum\limits_{r = 1}^R {{l_{r,k}}{\bf{F}}_{r,k}^H} {\bf{\Theta }}_r^H{\bf{G}}_{b,r}} \right)} } {{\bf{w}}_{b,j}}{s_{j}} + {n_k} ,\\
{{\hat y}}_k^e = \sum\limits_{b = 1}^B {\sum\limits_{j = 1}^K {\left( {{\bf{H}}_{b,e}^H + \sum\limits_{r = 1}^R {{l_{r,k}}{\bf{F}}_{r,e,k}^H} {\bf{\Theta }}_r^H{\bf{G}}_{b,r}} \right)} } {{\bf{w}}_{b,j}}{s_j} + {n_e}.
\end{align}
\end{subequations}

The virtual achievable secret rate of user $k$ can be written as
\begin{equation}\label{eq:35}
{{\hat R}_{s,k}} = {\left[{\kern 1pt} {\ln (1 + {{\hat \gamma }_k}) - \ln (1 + {{\hat \gamma }_{e,k}})}{\kern 1pt} \right]^ + },
\end{equation}
where the virtual SINRs of ${{\hat \gamma }_k}$ and ${{\hat \gamma }_{e,k}}$ are the same as (7a) and (7b), respectively, which will be given in the following sub-section. We introduce a parameter ${R_{{\rm{assign}}}}$ to indicate the maximum number of RISs served by each user, where ${R_{{\rm{assign}}}} \le R$. Therefore, the maximum WSSR problem with assigning RIS is formulated as
\begin{subequations}\label{eq:p36}
\begin{align}
{{\cal P}_{assign}}:&\mathop {\max }\limits_{{\textbf{L}_{k}},{\bf{W}},\boldsymbol{\mu} } {\rm{  }}  {{\hat R}_S} = \sum\limits_{k = 1}^K {{\eta _k}\left( {\ln \left( {1 + {{\hat \gamma }_k}} \right) - \ln \left( {1 + \hat \gamma _k^e} \right)} \right)} \\
{\rm{s}}{\rm{.t.}}& \ \ \  C1:\sum\limits_{k = 1}^K {{\rm{ }}{{\left\| {{{\bf{w}}_{b,k}}} \right\|}^2}}  \le {P_b}{\rm{, }}\forall b \in {\cal B},\\
&\ \ \ C2:{\rm{ }}\left| {{\theta _{r,n}}} \right| = 1,{\rm{ }}\forall r \in {\cal R},\forall n \in {\cal N},\\
&\ \ \ C4: {l_{r,k}} \in \left\{ {0,1} \right\},\forall r \in {\cal R},\forall k \in {\cal K},\\
&\ \ \ C5: \sum\limits_{r = 1}^R {{\kern 1pt} {l_{r,k}} \le } {\kern 1pt} {\kern 1pt} {R_{assign}},\forall k \in {\cal K}.
\end{align}
\end{subequations}

Due to the nonconvex objective function (36a), it is obvious that ${{\cal P}_{assign}}$ is a  nonconvex MINP, which is difficult to solve directively. Next, an effective algorithm is proposed to deal with it.

\subsection{Joint Optimization Algorithm for Assigning RIS}
When the variable ${l_{r,k}}$ is fixed, we only need to multiply both ${{\bf{F}}_{r,k}}$ and ${{\bf{F}}_{r,e,k}}$ by a coefficient ${l_{r,k}}$ to replace the original ${{\bf{F}}_{r,k}}$ and ${{\bf{F}}_{r,e,k}}$, and then the problem ${{\cal P}_{assign}}$ can be transformed into the same form as the problem ${{\cal P}^o}$ that can be solve by the proposed \textbf{Algorithm 1} proposed in Section III.

In this section, we mainly focus on the problem of optimizing ${\textbf{{L}}_{k}}$ when ${\bf{W}}$ and $\boldsymbol{\mu}$ are fixed. To transform ${{\cal P}_{assign}}$ into a solvable convex problem, we rewrite the mathematical form of the channel.
The reflection channels of the $k$-th user and the Eve eavesdropping on the $k$-th user are denoted as

\begin{algorithm}
	\renewcommand{\algorithmicrequire}{\textbf{Input:}}
	\renewcommand{\algorithmicensure}{\textbf{Output:}}
    \caption{Proposed Joint Optimization AO Algorithm.}
    \label{Algorithm2}
    \begin{algorithmic}[1]
        \REQUIRE All channels ${{\bf{H}}_{b,k}}$, ${{\bf{H}}_{b,e}}$, ${{\bf{F}}_{r,k}}$, ${{\bf{F}}_{r,e,k}}$, ${{\bf{G}}_{b,r}}$, and noise ${\sigma _k}$, ${\sigma _e}$.
        \ENSURE BF matrix ${\bf{W}}$, phase shift vector $\boldsymbol{\mu}$, User-RIS assigned vector $\textbf{{L}}_k$ and the virtual maximum WSSR ${{\hat R}_s}$.
        \STATE Initialize ${\bf{W}}^t$, ${\boldsymbol{\mu}}^t$, $\textbf{{L}}_k^t$, ${{\hat R}_s^t}$, $t=0$ and threshold $\varepsilon$;
        \WHILE{${R_s^t}-{R_{s}^{t-1}}>{\varepsilon}$}
        \STATE $t=t+1$;
        \STATE Update ${\bf{W}}^t$ by ${{\cal P}^1}$;
        \STATE Update $\boldsymbol{\mu}^t$ by ${{\cal P}^5}$;
        \STATE Update $\textbf{{L}}_k^t$ by ${{\hat {\cal P}}_{assign}}$;
        \STATE Update ${{\hat R}_s^t}$ by (36a);
        \ENDWHILE
        \STATE \textbf{return} ${\bf{W}}^t$, ${\boldsymbol{\mu}}^t$, $\textbf{{L}}_k^t$, and ${{\hat R}_s^t}$.
    \end{algorithmic}
\end{algorithm}

\begin{subequations}\label{eq:p37}
\begin{align}
& \sum\limits_{b = 1}^B {\sum\limits_{r = 1}^R {{l_{r,k}}{\bf{F}}_{r,k}^H} {\bf{\Theta }}_r^H{{\bf{G}}_{b,r}}} {{\bf{w}}_{b,k}} = \sum\limits_{r = 1}^R {{l_{r,k}}{\textbf{{a}}_{r,k}}}{{\bf{w}}_k}  = \textbf{{L}}_k^T{\textbf{{a}}_k}{{\bf{w}}_k},\\
& \sum\limits_{b = 1}^B {\sum\limits_{r = 1}^R {{l_{r,k}}{\bf{F}}_{r,e,k}^H} {\bf{\Theta }}_r^H{{\bf{G}}_{b,r}}} {{\bf{w}}_{b,k}} = \sum\limits_{r = 1}^R {{l_{r,k}}{\textbf{{a}}_{r,e,k}}}{{\bf{w}}_k}  = \textbf{{L}}_k^T{\textbf{{a}}_{e,k}}{{\bf{w}}_k},
\end{align}
\end{subequations}
where ${\textbf{{a}}_{r,k}} = {{\bf{F}}_{r,k}^H} {\bf{\Theta }}_r^H{{\bf{G}}_{r}}$, ${\textbf{{a}}_{r,e,k}} = {{\bf{F}}_{r,e,k}^H} {\bf{\Theta }}_r^H{{\bf{G}}_{r}}$, ${{\bf{G}}_r} = \left[ {{{\bf{G}}_{1,r}}, \ldots ,{{\bf{G}}_{b,r}}} \right]$, ${\textbf{{a}}_k} = {\left[ {\textbf{{a}}_{1,k}^T, \ldots ,\textbf{{a}}_{R,k}^T} \right]^T}$, and ${\textbf{{a}}_{e,k}} = {\left[ {\textbf{{a}}_{1,e,k}^T, \ldots ,\textbf{{a}}_{R,e,k}^T} \right]^T}$. For convenience, we define the following
\begin{subequations}\label{eq:p38}
\begin{align}
& \sum\limits_{b = 1}^B {\sum\limits_{r = 1}^R {\left( {{\bf{H}}_{b,k}^H + {l_{r,k}}{\bf{F}}_{r,k}^H{\bf{\Theta }}_r^H{{\bf{G}}_{b,r}}} \right)} } {{\bf{w}}_{b,k}} = \textbf{{u}}_k^H{\textbf{{b}}_k}{{\bf{w}}_k},\\
& \sum\limits_{b = 1}^B {\sum\limits_{r = 1}^R {\left( {{\bf{H}}_{b,e}^H + {l_{r,k}}{\bf{F}}_{r,e,k}^H{\bf{\Theta }}_r^H{{\bf{G}}_{b,r}}} \right)} } {{\bf{w}}_{b,k}} = \textbf{{u}}_k^H{\textbf{{b}}_{e,k}}{{\bf{w}}_k},
\end{align}
\end{subequations}
where ${\textbf{{u}}_k} = {\left[ {\textbf{{L}}_k^T,1} \right]^T}$, ${\textbf{{b}}_k} = \left[ {\textbf{{a}}_k^T,{{\bf{H}}_{d,k}}} \right]^T$, ${\textbf{{b}}_{e,k}} = \left[ {\textbf{{a}}_{e,k}^T,{{\bf{H}}_{d,e}}} \right]^T$, ${{\bf{H}}_{d,k}} = {\left[ {{\bf{H}}_{1,k}^T, \ldots ,{\bf{H}}_{b,k}^T} \right]^T}$, and ${{\bf{H}}_{d,e}} = {\left[ {{\bf{H}}_{1,e}^T, \ldots ,{\bf{H}}_{b,e}^T} \right]^T}$.
Then, the virtual SINR of the received signal of the $k$-th user and the Eve eavesdropping the $k$-th user's information can be calculated separately as
\begin{subequations}\label{eq:p39}
\begin{align}
& {{\hat \gamma }_k} = \frac{{{{\left| {\textbf{{u}}_k^H{\tilde{\textbf{{b}}}_k}{{\bf{w}}_k}} \right|}^2}}}{{\sum\limits_{j = 1,j \ne k}^K {{{\left| {\textbf{{u}}_k^H{\tilde{\textbf{{b}}}_k}{{\bf{w}}_j}} \right|}^2} + 1} }}, \\
& \hat \gamma _k^e = \frac{{{{\left| {\textbf{{u}}_k^H{\tilde{\textbf{{b}}}_{e,k}}{{\bf{w}}_k}} \right|}^2}}}{{\sum\limits_{j = 1,j \ne k}^K {{{\left| {\textbf{{u}}_k^H{\tilde{\textbf{{b}}}_{e,k}}{{\bf{w}}_j}} \right|}^2} + 1} }},
\end{align}
\end{subequations}
where ${\tilde{\textbf{{b}}}_k}={\sigma _k^{-1}}{\textbf{{b}}_k}$ and ${\tilde{\textbf{{b}}}_{e,k}}={\sigma _e^{-1}}{\textbf{{b}}_{e,k}}$.
Thus, the achievable virtual secery rate from BSs to the $k$-th user can be expressed as
\begin{equation}\label{eq:40}
\begin{split}
& \ \ \ \ \ \ln \left( {1 + {{\hat \gamma }_k}} \right) - \ln \left( {1 + \hat \gamma _k^e} \right)\\
 = & \ln \left( {\frac{{\sum\limits_{j = 1}^K {{{\left| {\textbf{{u}}_k^H{\tilde{\textbf{{b}}}_k}{{\bf{w}}_j}} \right|}^2} + 1} }}{{\sum\limits_{j = 1,j \ne k}^K {{{\left| {\textbf{{u}}_k^H{\tilde{\textbf{{b}}}_k}{{\bf{w}}_j}} \right|}^2} + 1} }}\frac{{\sum\limits_{j = 1,j \ne k}^K {{{\left| {\textbf{{u}}_k^H{\tilde{\textbf{{b}}}_{e,k}}{{\bf{w}}_j}} \right|}^2} + 1} }}{{\sum\limits_{j = 1}^K {{{\left| {\textbf{{u}}_k^H{\tilde{\textbf{{b}}}_{e,k}}{{\bf{w}}_j}} \right|}^2} + 1} }}} \right)\\
 =& \ln \left( {\frac{{\sum\limits_{j = 1}^K {{\rm{Tr}}\left( {{{\bf{D}}_{k,j}}{{\bf{U}}_k}} \right) + 1} }}{{\sum\limits_{j = 1,j \ne k}^K {{\rm{Tr}}\left( {{{\bf{D}}_{k,j}}{{\bf{U}}_k}} \right) + 1} }}\frac{{\sum\limits_{j = 1,j \ne k}^K {{\rm{Tr}}\left( {{{\bf{D}}_{e,k,j}}{{\bf{U}}_k}} \right) + 1} }}{{\sum\limits_{j = 1}^K {{\rm{Tr}}({{\bf{D}}_{e,k,j}}{{\bf{U}}_k}) + 1} }}} \right)\\
 = &\ \ \  \  T_k^1 + T_k^2 - T_k^3 - T_k^4,
\end{split}
\end{equation}
where ${{\bf{U}}_k} = {\textbf{{u}}_k}\textbf{{u}}_k^H$ , ${{\bf{D}}_{k,j}} = {\textbf{{b}}_k}{{\bf{w}}_j}{\bf{w}}_j^H\textbf{{b}}_k^H$, ${{\bf{D}}_{e,k,j}} = {\textbf{{b}}_{e,k}}{{\bf{w}}_j}{\bf{w}}_j^H\textbf{{b}}_{e,k}^H$, and $T_k^1$, $T_k^2$, $T_k^3$, $T_k^4$ represented by:
\begin{subequations}\label{eq:p41}
\begin{align}
&T_k^1 = \ln \left( {\sum\limits_{j = 1}^K {{\rm{Tr}}\left( {{{\bf{D}}_{k,j}}{{\bf{U}}_k}} \right) + 1} } \right){\kern 1pt} {\kern 1pt} ,\\
&T_k^2 = \ln \left( {\sum\limits_{j = 1,j \ne k}^K {{\rm{Tr}}\left( {{{\bf{D}}_{e,k,j}}{{\bf{U}}_k}} \right) + 1} } \right),\\
&T_k^3 = \ln \left( {\sum\limits_{j = 1,j \ne k}^K {{\rm{Tr}}\left( {{{\bf{D}}_{k,j}}{{\bf{U}}_k}} \right) + 1} } \right),\\
&T_k^4 = \ln \left( {\sum\limits_{j = 1}^K {{\rm{Tr}}({{\bf{D}}_{e,k,j}}{{\bf{U}}_k}) + 1} } \right).
\end{align}
\end{subequations}
It is clear that (40) is a difference of convex (DC) function, rather than convex.
In order to convert (40) into a convex function, we introduce the first-order Taylor to approximate the global upper bounds of $T_k^3$ and $T_k^4$. First of all, for the convex function $f(b)$ we have:
\begin{subequations}\label{eq:p42}
\begin{align}
& \left( {1 - \lambda } \right)f\left( a \right) + \lambda f\left( b \right) \le f\left( {\left( {1 - \lambda } \right) + \lambda b} \right),\\
& f\left( b \right) \le f\left( a \right) + \frac{{f\left( {a + \lambda \left( {b - a} \right)} \right) - f\left( a \right)}}{\lambda }    \nonumber \\
&\qquad =f\left( a \right) + \nabla f\left( a \right)\left( {b - a} \right),\lambda  \to 0.
\end{align}
\end{subequations}
Therefore, around the fixed point $\left\{ {{\mathbf{U}}_k^t} \right\}$ we can get
\begin{subequations}\label{eq:p43}
\begin{align}
 T_k^3\left( {{{\bf{U}}_k}} \right) & \le T_k^3\left( {{\bf{U}}_k^t} \right) + {\rm{Tr}}\left( {{\nabla _{{{\bf{U}}_k}}}T_k^3{{\left( {{\bf{U}}_k^t} \right)}^H}\left( {{{\bf{U}}_k} - {\bf{U}}_k^t} \right)} \right)  \nonumber  \\
&= \tilde T_k^3\left( {{{\bf{U}}_k},{\bf{U}}_k^t} \right) ,\\
 T_k^4\left( {{{\bf{U}}_k}} \right) & \le T_k^4\left( {{\bf{U}}_k^t} \right) + {\rm{Tr}}\left( {{\nabla _{{{\bf{U}}_k}}}T_k^4{{\left( {{\bf{U}}_k^t} \right)}^H}\left( {{{\bf{U}}_k} - {\bf{U}}_k^t} \right)} \right)   \nonumber  \\
&= \tilde T_k^4\left( {{{\bf{U}}_k},{\bf{U}}_k^t} \right) ,
\end{align}
\end{subequations}
where
\begin{subequations}\label{eq:p44}
\begin{align}
{\nabla _{{{\bf{U}}_k}}}T_k^3\left( {{\bf{U}}_k^t} \right) &= \frac{{\sum\limits_{j = 1,j \ne k}^K {{{\bf{D}}_{k,j}}} }}{{\sum\limits_{j = 1,j \ne k}^K {{\rm{Tr}}\left( {{{\bf{D}}_{k,j}}{\bf{U}}_k^t} \right) + 1} }} ,\\
{\nabla _{{{\bf{U}}_k}}}T_k^4\left( {{\bf{U}}_k^t} \right) &= \frac{{\sum\limits_{j = 1}^K {{{\bf{D}}_{e,k,j}}} }}{{\sum\limits_{j = 1}^K {{\rm{Tr}}\left( {{{\bf{D}}_{e,k,j}}{\bf{U}}_k^t} \right) + 1} }} .
\end{align}
\end{subequations}
Define $g\left( {{{\bf{U}}_k}} \right) =  - T_k^1 - T_k^2 + T_k^3 + T_k^4$ and $\tilde g\left( {{{\bf{U}}_k}} \right) =  - T_k^1 - T_k^2 + \tilde T_k^3 + \tilde T_k^4$. As long as $g\left( {{{\bf{U}}_k}} \right) < 0$ and $\tilde g\left( {{{\bf{U}}_k}} \right) \le 0$, $g\left( {{{\bf{U}}_k}} \right) \le \tilde g\left( {{{\bf{U}}_k}} \right)$ are satisfied.

Therefore, after fixing ${\bf{W}}$ and $\boldsymbol{\mu}$, the problem ${{\cal P}_{assign}}$ can be transformed into a solvable problem as follows:
\begin{subequations}\label{eq:p45}
\begin{align}
 {{\tilde {\cal P}}_{assign}} :& \mathop {\min {\kern 1pt} }\limits_{{{\mathbf{U}}_k}} \  \sum\limits_{k = 1}^K {{\eta _k}} \tilde g\left( {{{\mathbf{U}}_k}} \right) \\
 {\rm{s}}{\rm{.t.}} &\ \ C4: {l_{r,k}} \in \left\{ {0,1} \right\},\forall r \in {\cal R},\forall k \in {\cal K},\\
&\ \ C5: \sum\limits_{r = 1}^R {{\kern 1pt} {l_{r,k}} \le } {\kern 1pt} {\kern 1pt} {R_{assign}},\forall k \in {\cal K},\\
  &\ \ C6: {{\mathbf{U}}_k} \succeq {\mathbf{0}},\forall k \in {\cal K}.
\end{align}
\end{subequations}
Due to the constraints of $C4$ and $C5$, problem ${{\tilde {\cal P}}_{assign}}$ is a quadratic problem of zero-one programming, which is still difficult to solve. Here, we use LCR [33] relaxation constraints to obtain a suboptimal solution for the objective function. Accordingly, ${{\tilde {\cal P}}_{assign}}$ can be rewritten as:
\begin{subequations}\label{eq:p46}
\begin{align}
& {{\hat {\cal P}}_{assign}} : \mathop {\min }\limits_{{\textbf{{u}}_k},{{\bf{U}}_k}} {\kern 1pt} {\kern 1pt} \sum\limits_{k = 1}^K {{\eta _k}} \tilde g\left( {{{\mathbf{U}}_k}} \right)\\
&\ {\rm{s}}{\rm{.t.}} \ C7: {\kern 1pt} {\left[ {{{\bf{U}}_k}} \right]_{i,i}} =  {\left[ {{\textbf{{u}}_k}} \right]_i},\forall i \in \left\{ {1, \ldots ,R + 1} \right\},\forall k \in {\cal K},\\
 &\qquad  C8: {\kern 1pt}{[{\textbf{{u}}_k}]_{R + 1}} = 1,\forall k \in {\cal K},\\
 &\qquad  C9:{\kern 1pt} {\bf{1}}_{R + 1}^T{\textbf{{u}}_k} \le {R_{assign}} + 1,\forall k \in {\cal K},\\
 &\qquad  C10:{\kern 1pt} {\bf{1}}_{R + 1}^T{{\bf{U}}_k} \le \left( {{R_{assign}} + 1} \right)\textbf{{u}}_k^T,\forall k \in {\cal K},\\
 &\qquad  C11: {\kern 1pt} \left[ {\begin{array}{*{20}{c}}
1&{\textbf{{u}}_k^T}\\
{{\textbf{{u}}_k}}&{{{\bf{U}}_k}}
\end{array}} \right]{\kern 1pt}  \succeq {\bf{0}},\forall k \in {\cal K},
\end{align}
\end{subequations}
Here the constraints become convex, and ${{\hat {\cal P}}_{assign}}$ becomes a resoluble semidefinite programming (SDP) problem. The optimal solution for ${\textbf{{u}}_k}$ in problem ${{\hat {\cal P}}_{assign}}$ can be easily obtained according to modern SDP solvers. Afterwards, to restore to discrete $\textbf{{L}}_k$, we set the first ${R_{assign}} + 1$ larger items of ${\textbf{{u}}_k}$ to be 1, and the rest are 0, and then take the first $R$ items of ${\textbf{{u}}_k}$ as $\textbf{{L}}_k$.

Finally, we summarize the scheme for solving the original problem ${{\cal P}_{assign}}$ as \textbf{Algorithm 2}. Specifically, by fixing other variables, the solving sub-problems ${{{\cal P}}_{1}}$, ${{{\cal P}}_{5}}$ and ${{\hat {\cal P}}_{assign}}$ are updated separately through the AO approach until the WSSR converges.
\section{Supplementary Framework}
Due to the constraints of realistic conditions, it is challenging to achieve continuous phase adjustment for the reflection coefficient (RC). In this section, we mainly discuss the mode of RC discrete phase adjustment under non-ideal conditions, and analyze the complexity of the above two algorithms.

\subsection{Non-Ideal Discrete Phase Shift}
In Sub-Section II-B, we define ${\cal F}$ as an ideal RIS's RC that can be continuously adjusted. However, due to the constraints of hardware, the phase of the RIS elements are generally discrete. Here, we still define the amplitude coefficient ${\kappa _{r,n}}$ as constant 1, and re-define the ideal RC case as ${\cal F}_1$ and the discrete RC case as ${\cal F}_2$ [34], [35], i.e.,
\begin{subequations}\label{eq:p47}
\begin{align}
&{{\cal F}_1}{\rm{ = }}\left\{ {{\theta _{r,{\rm{n}}}}\left| {{\theta _{r,{\rm{n}}}}{\rm{ = }}{e^{j{\varphi _{r,n}}}}} \right.,{\varphi _{r,n}} \in \left[ {0,\left. {2\pi } \right)} \right.} \right\},\\
&{{\cal F}_2}{\rm{ = }}\left\{ {{\theta _{r,{\rm{n}}}}\left| {{\theta _{r,{\rm{n}}}}{\rm{ = }}{e^{j{\varphi _{r,n}}}}} \right.,{\varphi _{r,n}} \in \left\{ {0,\frac{{2\pi }}{L}, \ldots ,\frac{{2\pi \left( {L - 1} \right)}}{L}} \right\}} \right\},
\end{align}
\end{subequations}
where $L$ means ${{\cal F}_2}$ has $L$ discrete phase shift values.

To solve the discrete RC problem, we can project continuous ${\theta _{r,{\rm{n}}}}$ from ${{\cal F}_1}$ to ${{\cal F}_2}$ to obtain discrete ${\theta _{r,{\rm{n}}}^o}$ [36]. Specifically, we relax the discrete RC ${\theta _{r,{\rm{n}}}^o}$ to a continuous RC ${\theta _{r,{\rm{n}}}}$, and then obtain the sub-optimal solution ${\boldsymbol{\mu} ^*}$ in each iteration to optimize the continuous ${\boldsymbol{\mu}}$, thereby obtaining the sub-optimal $\theta _{r,n}^*$. Project $\theta _{r,n}^*$ from ${{\cal F}_1}$ to ${{\cal F}_2}$, the specific method of projection is
$\angle {\theta _{r,{\rm{n}}}^o} = \mathop {\arg {\rm{ min}}}\limits_{\Phi  \in {{\cal F}_2}} \left| {\angle \theta _{r,n}^* - \angle \Phi } \right|.$
Therefore, we can obtain the sub-optimal solution of the phase shift optimization for each iteration sub-problem. Although the previous joint optimization is based on continuous phase shift, the convergence is not yet known. The simulation in Section VI show that the discrete phase shift optimization also has good convergence.

\subsection{Algorithmic Complexity}
In this subsection, we calculate the computational complexity of the proposed two algorithms. When applying the ADMM algorithm to solve $\boldsymbol{\mu}$, the computational complexity is ${\cal O}( {{T_{ADMM}}{{( {RN + 1} )}^2}} )$, where ${T_{ADMM}}$ symbolize the number of iterations required for $\boldsymbol{\mu}$ to converge. Thus, the computational complexity of \textbf{Algorithm 1} is  ${\cal O} ( {T_o}( {B^2}{M^2}{K^2}( {NR + 1} ) + BM{K^2}{{( {NR + 1} )}^2} + {T_{ADMM}} $
${{( {RN + 1} )}^2} ) )$, where ${T_o}$ is the number of iterations required for algorithm convergence. Similarly, the computational complexity of \textbf{Algorithm 2} is
${\cal O} \left( {T_o}\left( {B^2}{M^2}{K^2}( {NR + 1} ) \right. \right.$ $\left. \left.+ BM{K^2}{{( {NR + 1} )}^2} + {T_{ADMM}} {{( {RN + 1} )}^2} +K{{( {R + 1} )}^3} \right. \right.$ 
$ {{( {KR + 1} )}^{2.5}}) )$.
\section{Numerical Results}
In this section, we numerically analyze the performance of the proposed AO scheme and the extended assigning RIS scheme.
\begin{figure}[t]
  \centering
  \includegraphics[scale = 0.3]{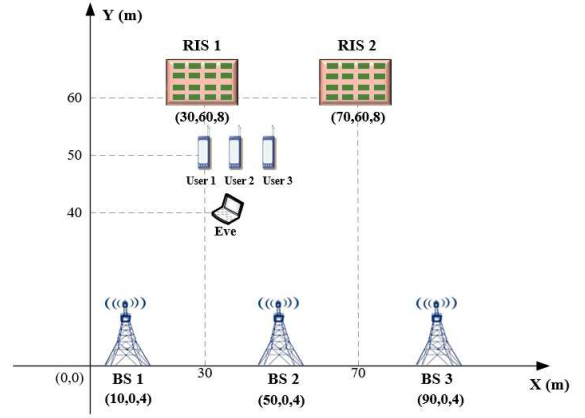}
  \caption{The IRS-aided cell-free network MISO system model.}
  \label{fig.3}
\end{figure}
We use three-dimensional coordinates in the evaluation to demonstrate the simulation scene as shown in Fig. 3. We deploy three BSs to serve three users and one Eve. Meanwhile, we assume that there are buildings that affect users and Eve to receive signals, and thus we deploy two RISs on the distant high ground to enhance the secrecy rate. Assume that three BSs have a height of 4 m and are located at (10 m, 0 m, 4 m), (50 m, 0 m, 4 m) and (50 m, 0 m, 4 m). The three users and Eve in the distance are located at (30 m, 50 m, 1.5 m), (35 m, 50 m, 1.5 m), (40 m, 50 m, 1.5 m), and (35 m, 40 m, 1.5 m) respectively, and the two RISs farther away are located at (30 m, 60 m, 8 m) and (70 m, 60 m, 8 m), respectively.

To match the real scene in the cell-free network, we set a small number of antennas in the BSs to $M=5$, and the transmit power of the BS to $P_b=0$ dBm. The noise power of user $k$ and Eve is $\sigma _k^2=\sigma _e^2=-80$ dBm, and the weight of the user is set to ${\eta _k}=1$ for ease of calculation. Assume that the number of elements in each RIS is $N=50$. Both our direct channel and reflected channel follow a distance-dependent path loss model [26]. We formulate the channel fading model as
 ${\bf{H}}{\rm{ = }}\sqrt {{L_0}{{\left( {\frac{d}{{{d_0}}}} \right)}^{ - \tau }}} {{\bf{H}}^ * },$
 where ${L_0}$ represents the path loss of the reference distance ${d_0}=1$ m, ${d}$ represents the distance between the transmitting place and the receiving place, $\tau$  represents the path loss exponent, and ${{\bf{H}}^ * }$ represents the small-scale fading model. Here we consider ${{\bf{H}}^ * }$ as a Rician fading channel [30], which is calculated as
  ${{{\bf{H}}^ * } = \sqrt {\frac{{K'}}{{K' + 1}}} {\bf{H}}_{LoS}^ *  + \sqrt {\frac{1}{{K' + 1}}} {\bf{H}}_{NLoS}^ *},$
  \begin{table}[t]
	\caption{List of Key Notations}
	\label{KeyNotations}
	\centering
	\begin{tabular}{p{75pt}|p{140pt}}
		\hline
        Parameter & Value \\\hline
		Path loss exponent & ${\tau _{BU}} = {\tau _{BE}} = 3.5$, ${\tau _{RU}} = {\tau _{RE}} = 2.5$, ${\tau _{BR}} = 2.0$   \\\hline
		Rician channel factor & ${{K'}_{BU}} = {{K'}_{BE}} = 0$, ${{K'}_{BR}} = {{K'}_{RU}} = {{K'}_{RE}} = 3$ \\\hline
        The Inter-antenna separation distance  & ${d_r}={d_t}={\lambda}/2$ \\\hline
        Path loss at 1 meter & ${L_0} =  - 30$ dB  \\\hline
	\end{tabular}
 \end{table}
 where ${K'}$ is the Rician factor, and ${\bf{H}}_{LoS}^ *$  and ${\bf{H}}_{NLoS}^ *$ are the line-of-sight (LoS) and Rayleigh fading components, respectively. ${\bf{H}}_{LoS}^ *$  is calculated as ${\bf{H}}_{LoS}^ *  = {\bf{q}}\left( {{\vartheta ^{AoA}}} \right){\bf{q}}{\left( {{\vartheta ^{AoD}}} \right)^H}$, where ${\bf{q}}\left( {{\vartheta ^{AoA}}} \right) = \exp {\left( {0, \ldots ,j\frac{{2\pi {d_r}}}{\lambda }({D_r} - 1)\sin {\vartheta ^{AoA}}} \right)^T}$, and ${\bf{q}}\left( {{\vartheta ^{AoD}}} \right) = \exp {\left( {0, \ldots ,j\frac{{2\pi {d_t}}}{\lambda }({D_t} - 1)\sin {\vartheta ^{AoD}}} \right)^T}$, here ${\vartheta ^{AoA}}$, ${d_r}$, and ${D_r}$ respectively represent the angle of arrival (AoA) of the receiver, the number of antennas, and the separation distance between antennas. ${\vartheta ^{AoD}}$, ${d_r}$, and ${D_r}$ respectively represent the angle of departure (AoD) of the transmitter, the number of antennas, and the separation distance between antennas, and $\lambda$ is the wavelength [30]. The specific parameters setting can be found in $\text{Table I}$.

To show the convergence of our proposed algorithms, Fig. 4 plots the WSSR versus the number of iterations.
Here, the legends have the following meanings:
{\bf{Ideal RIS case}}: without assigning RIS and continuous phase shift and perfect CSI obtained from Algorithm 1.
{\bf{Discrete phase shift}}: discrete 3-bit phase shift while performing the steps mentioned in Section V-A.
{\bf{Assigning RIS scheme}}: assuming $R_{Assign}=1$, then execute Algorithm 2 with the phase shift being continuous.
{\bf{Random phase shift}}: the initial phase shift is random, and only the BF vector ${\bf{W}}$ is optimized without optimizing the phase shift.
{\bf{Without RIS}}: there is no RIS in the scene.

From Fig. 4, we can observe that all curves increase with the number of iterations until convergence, which is achieved within 13 iterations. This confirms that the relaxation constraint conditions of the ``Assigning RIS scheme'' 
have minor effect on the convergence. Although the ``Discrete phase shift scheme'' is previously unpredictable, it has now also been shown to converge quickly.
\begin{figure}[t]
  \centering
  \includegraphics[scale = 0.42]{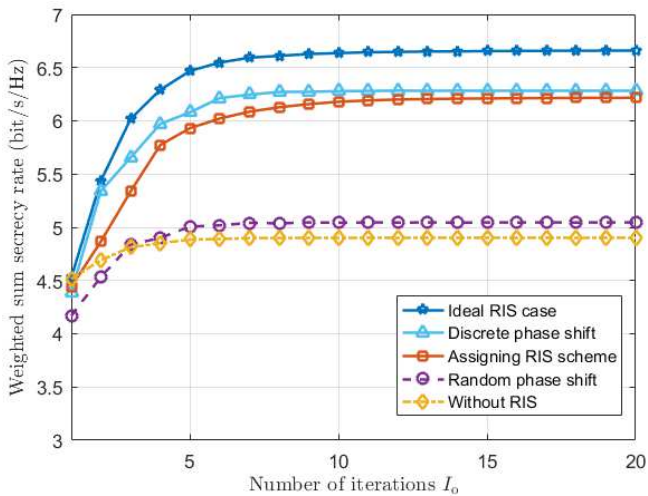}
  \caption{Weighted sum secrecy rate versus the number of iterations.}
  \label{fig.4}
\end{figure}
In addition, we can observe that the RIS-aided WSSR is much higher than the WSSR without RIS, which also proves the importance of the RISs in improving the secrecy rate. For the ``Assigning RIS scheme'', although the user is served by one RIS at most, its performance is close to that of the ``Ideal case''. One can also observe that the performance of ``Random phase shift'' without optimizing the phase shift is close to that of the network without RIS assistance, which shows that the ``Assigning RIS scheme'' is reasonable in numerical evaluation compared with the situation that all RISs serve users.

Next, we evaluate the impact of the key parameters on the WSSR. The parameters are the same as the above unless otherwise stated.

\emph{1) Relationship between the WSSR and the BS transmit power}: As shown in Fig. 5, the WSSR grow rapidly with the BS transmit power under all schemes. It is worth noting that the WSSR of our proposed ``Assigning RIS scheme'' is close to that of the ``Ideal RIS case''. Specifically, one can observe that when the BS transmit power is small, the WSSR of all schemes is close to zero, and as the transmit power increases, the WSSR ratio of all RIS-aided schemes relative to the without RIS-aided scheme gradually decreases. Here, it can be explained that when the BS transmit power is relatively small, the WSSR contribution of RIS to the network is more obvious, because the BSs tend to assign more power to the reflection links. When the BS transmit power is relatively high, the WSSR contribution of RIS to the network is significantly reduced, because the BSs are more inclined to assign more power to the direct channel for transmission. In addition, we can also find that the performance of the ``Discrete phase shift'' is relatively closer to the ``Without RIS'' as the transmit power increases.
\begin{figure}[t]
  \centering
  \includegraphics[scale = 0.43]{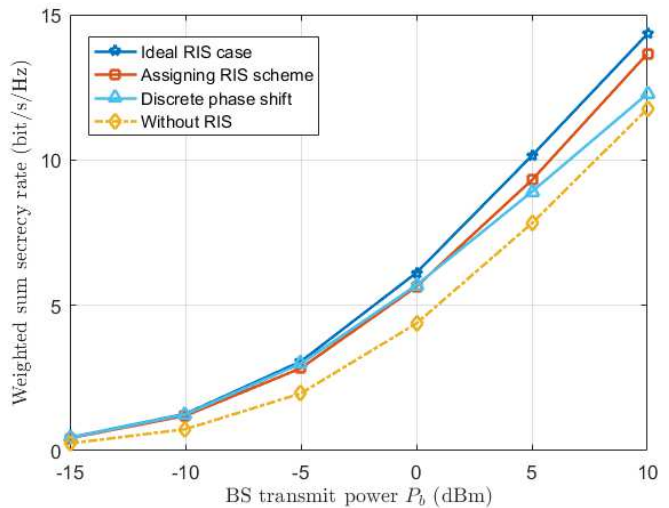}
  \caption{Weighted sum secrecy rate versus the BS transmit power $P_b$.}
  \label{fig.5}
  \end{figure}

\emph{2) Relationship between the WSSR and the number of RIS elements}: Fig. 6 shows the WSSR versus the number of RIS elements. As expected, the WSSR remains a constant for different number of RIS elements under Without RIS scheme. In contrast, the WSSR for all cases with RIS increases approximately linearly with the RIS elements.  This can be explained as follows:
 as $N$ increases, the RIS can receive and reflect more signal energy to the user, yielding a stronger received signals. Although the RIS with larger elements can obtain a higher WSSR, it requires more overhead to obtain the CSI. Thus, we need to balance the system performance and overhead, and then select the suitable number of RIS elements in practice.
\begin{figure}[t]
  \centering
  \includegraphics[scale = 0.42]{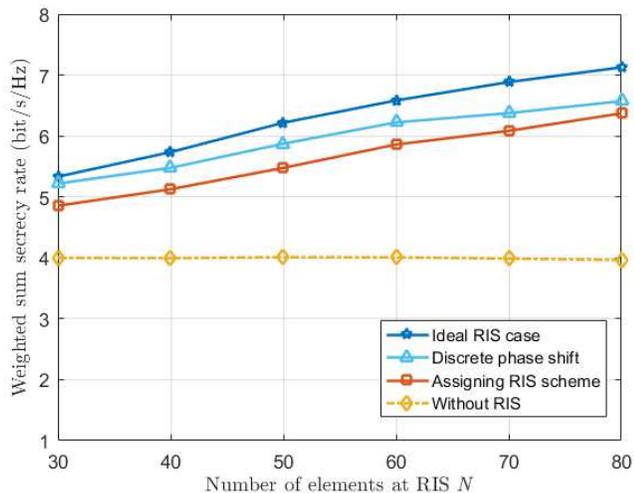}
  \caption{Weighted sum secrecy rate versus the number of RIS elements $N$.}
  \label{fig.6}
\end{figure}

\emph{3) Relationship between the WSSR and the distance $L_0$}: Here we set the two-dimensional coordinates of three users and the Eve to (${L_0}-5$ m, 50 m), ($L_0$ m, 50 m), ($L_0+5$ m, 50 m), and ($L_0$ m, 40 m) respectively, and plot Fig. 7 according to the relationship between the WSSR and $L_0$. It can be observed that the WSSR reaches its peak at $L_0=30$ m and $L_0=70$ m with the RIS-aided curve. That is, as the user and Eve move away from these two points, the WSSR decreases. It indicates that when users and Eve approach the RIS, the signal reflected by the RIS will be stronger, so the users can obtain a higher secrecy rate. Based on the above, in practice we can obtain a better security performance through reasonable deployment of RISs. In addition, we can also observe that the average performance loss of the ``Assign RIS scheme'' compared to the ``Ideal RIS case'' is minor, which also shows the effectiveness of our proposed ``Assign RIS scheme''.
\begin{figure}[t]
  \centering
  \includegraphics[scale = 0.42]{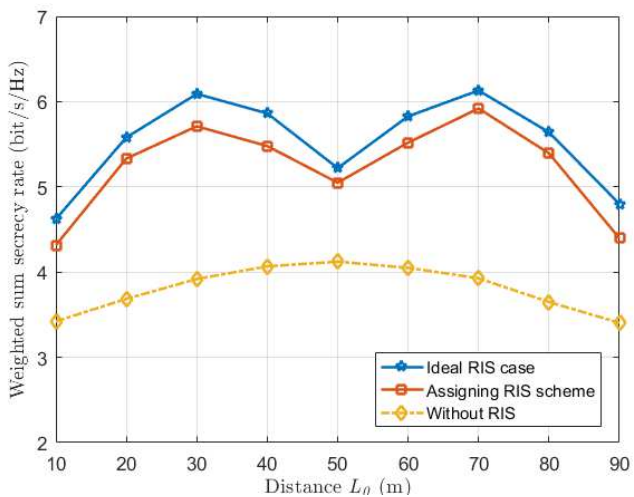}
  \caption{Weighted sum secrecy rate versus the distance $L_0$.}
  \label{fig.7}
\end{figure}

\begin{figure}[t]
  \centering
  \includegraphics[scale = 0.42]{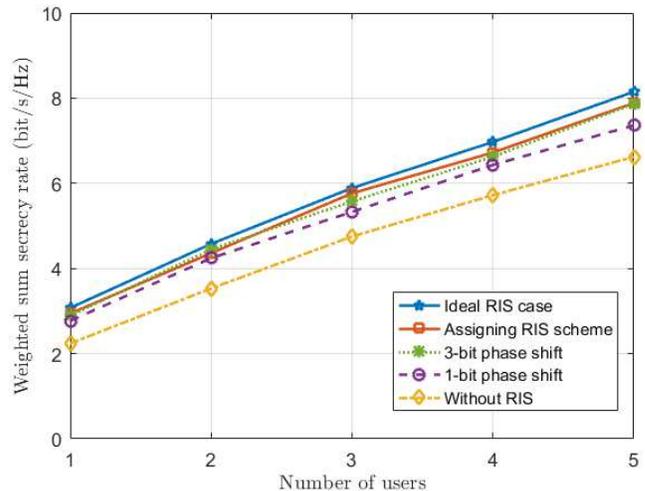}
  \caption{Weighted sum secrecy rate versus the number of users.}
  \label{fig.8}
\end{figure}

\emph{4) Relationship between the WSSR and the number of users}: Fig. 8 illustrates the WSSR versus the number of users. Here, we reset the two-dimensional coordinate position of the $k$-th user to (30+$k$ m, 50 m), and the total number of users to $K$. We also redefine 2 curves (``1-bit phase shift'' means 1-bit discrete phase shift scheme, and ``3-bit phase shift'' means 3-bit discrete phase shift scheme). Generally, we can observe that the WSSR increases with the number of users.
 Here, we can explain the above phenomenon as follow:  when the number of users is less than that of BS antennas, the interference among users has little effect on the secrecy rate. Therefore, the WSSR increases with the number of users.
In addition, we can also observe that as the resolution of the phase shifter increases, the performance of the system also increases. Nonetheless, it requires higher technical complexity and overhead. Finally, we can find that the solution ``Assign RIS scheme'' can still maintain similar performance to the ``Ideal RIS case'', which again indicates the effectiveness of our proposed  \textbf{Algorithm 2}.

\emph{5) Reconstruction of the relationship between the WSSR and the number of RIS elements}: To evaluate the impact of the number of RISs allocated by Algorithm 2 on the overall performance of the system, in this part, we re-plan the system model of the network. Specifically, we added the number of RISs to $R=5$ and located in ($10+20(r-1)$ m, 60 m), and further define three schemes $R_{assign}=1$, $R_{assign}=2$, and $R_{assign}=3$ respectively, and other system parameters are assumed to be the same as VI-A. On this basis, we plot Fig. 9 to show the variation of WSSR over the number of RIS elements.

\begin{figure}[t]
  \centering
  \includegraphics[scale = 0.42]{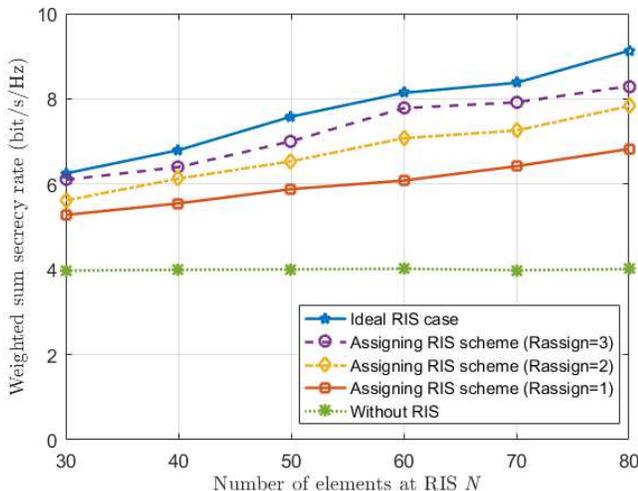}
  \caption{Weighted sum secrecy rate versus the number of elements at RIS.}
  \label{fig.9}
\end{figure}

From Fig. 9, we intuitively find that as the number of arranged RISs increases, the WSSR with the RIS-aided scheme grows much more than that in Fig. 6. Meanwhile, the WSSR increases with the RIS elements, which is easy to understand. Additionally, we can find that as the number of RISs allocated to each user increases, the WSSR also increases. It is because that as the number of RISs increases, the signals reflected by RISs becomes stronger, and thus users can obtain a higher secrecy rate.
Although users with more allocated RISs can achieve a higher WSSR, it requires more overhead to obtain the CSI, increasing the technique complexity. Therefore, we need to balance system performance and overhand, and select the suitable number of RISs services for users in practice.
\section{Conclusion}
In this paper, we investigated the secure transmission in the RIS-aided cell-free networks. We proposed a joint active and passive optimization at the BSs and RISs to maximize the WSSR. To lower the complexity and overhead, we extended the proposed scheme to the cases with discrete phase shift and RIS nulling. Simulation results showed that the distributed low cost RIS-aided cell free network can obtain a better performance than the traditional cell-free networks. Meanwhile, it was illustrated that there exists a tradeoff between system performance and complexity/overhead.

\end{document}